\documentclass[preprint,aps,groupedaddress]{revtex4}
\usepackage{graphicx}
\usepackage{amsmath,amssymb,amsthm}
\usepackage{bm}

\begin{document}
\title{Explosive magnetic reconnection caused by an X-shaped current-vortex
layer in a collisionless plasma}
\author{M.~Hirota, Y.~Hattori}
\affiliation{Tohoku University, Sendai, Miyagi 980-8677, Japan}

\author{P. J. Morrison}
\affiliation{Department of Physics and Institute for Fusion Studies\\
University of Texas at Austin, Austin, Texas 78712 USA}

\begin{abstract}

A mechanism for explosive magnetic reconnection is investigated by
analyzing the nonlinear evolution of a collisionless tearing mode in
a two-fluid model that includes the effects of electron  inertia and
temperature.  These effects cooperatively enable  a fast
reconnection by forming an X-shaped current-vortex layer centered at
the reconnection point.  A high-resolution simulation of this model for an 
unprecedentedly small electron skin depth $d_e$ and ion-sound gyroradius
$\rho_s$, satisfying $d_e=\rho_s$,  shows an explosive tendency  for 
nonlinear growth  of the tearing mode, where it is newly found that
the explosive widening of the X-shaped layer occurs locally around the
reconnection point with  the length of the X shape being shorter than the
domain length and the wavelength of the linear tearing mode.  The
reason for the onset of this locally enhanced reconnection is
explained theoretically by developing a novel nonlinear and  
nonequilibrium inner solution that  models the local X-shaped layer,  and
then matching it to an outer solution that is approximated by a linear
tearing eigenmode with a shorter wavelength than the domain length.  This
theoretical model proves that the local reconnection can release the
magnetic energy more efficiently than the global one and the estimated
scaling of the explosive growth rate agrees well with the simulation
results.
\end{abstract}

\maketitle

\section{Introduction}

Clarification of the mechanisms of fast magnetic reconnections in space and
laboratory plasmas is a fundamental issue that has been
tackled by plasma physicists for more than half a century~\cite{Sweet,Petschek}.
In particular, the explosive release  of magnetic energy, observed in solar flares,
magnetospheric substorms and tokamak sawtooth collapses, 
 indicates  that the magnetic reconnections might further accelerate through
nonlinear and nonequilibrium processes.
It is widely accepted that the resistive magnetohydrodynamic (MHD) model cannot reproduce 
such  fast reconnection 
unless a locally enhanced resistivity is artificially introduced~\cite{Ugai}.
Since the plasmas in these explosive events
are relatively collisionless, the resistivity is thought to be physically less important than
 non-collisional microscopic effects such
as electron inertia, Hall current, ion gyroradius effects, and so on, which are
all neglected in the classical MHD model.
Consequently, there has been considerable effort in recent years in studying a
variety of two-fluid~\cite{Hazeltine,Schep,Kuvshinov,Fitzpatrick},
gyro-fluid~\cite{Snyder,Waelbroeck} and gyrokinetic~\cite{Frieman,Zocco}
models to understand reconnection  in collisionless plasmas. Until now, many simulation
results~\cite{Aydemir,Ottaviani,Cafaro,Bhattacharjee,Matsumoto,Biancalani,Comisso,Ishizawa} have shown that   collisionless reconnection tends to accelerate into  a  nonlinear phase. However, the  
theoretical understanding of this process  is very limited.

The primary computational obstacle  is that
the nonlinear acceleration phase is observed only when 
the magnetic island width (or  the amplitude of the tearing
mode) exceeds the microscopic scales while  sufficiently smaller 
than the scale of the equilibrium magnetic shear. This suggests the importance of 
making the microscopic scales as small as possible, yet  affording very high spatial
resolution to prolong the acceleration phase.  Consequently, it has not been clear in previous simulations how 
reconnection is accelerated and whether or not it is explosive.

The nonlinear theory for explosive magnetic reconnection has remained elusive 
 because of the difficulty of solving  strongly nonlinear and nonequilibrium fluid motion with multiple
scales.  The method of asymptotic matching has been only applicable to
linear stability  of collisionless tearing modes, where the island
width is assumed to be much smaller than any microscopic
scale~\cite{Drake,Basu,Porcelli,Zocco}.
In the  dissipationless limit, recent studies take advantage of the
Hamiltonian structure \cite{morrison-greene,Hazeltine2,morrison98}  of the collisionless two-fluid models~\cite{Cafaro,Kuvshinov2,Grasso,Tassi,Comisso}. These studies  show
that the two-fluid effects distort the conservation laws (frozen-in variables) and hence
permit magnetic reconnections with  ideal fluid motion. 

In the presence of only electron inertia in the two-fluid model, an
ideal fluid motion develops an elongated current layer with Y-shaped
ends, where the layer width is comparable to the electron skin depth
($d_e$)~\cite{Ottaviani}.  On the other hand, even faster reconnections
due to the formations of X-shaped current-vortex layers are observed
numerically~\cite{Aydemir,Cafaro} when the effect of electron temperature is taken into
account and the ion-sound gyro-radius ($\rho_s$) is comparable to or
larger than $d_e$.  This distinction between the Y
and X shapes seems to be crucial in determining the reconnection speed,
in analogy with that between the Sweet-Parker~\cite{Sweet} and
Petschek~\cite{Petschek} models for resistive reconnections.  Although
several pioneering works~\cite{Ottaviani,Bhattacharjee} have attempted theoretical 
explanations   of  the explosive growth  of these nonlinear tearing modes, we note that their predictions are  not in quantitative agreement  with the high-resolution simulation results given in
Ref.~\cite{Hirota} and the present work.

The goal of this paper is to clarify an explosive mechanism for 
collisionless reconnection caused by the interplay of the
effects of electron inertia and temperature.
To this end, we analyze the  simple Hamiltonian two-fluid model given
explicitly in Sec.~\ref{sec:model}, both numerically and analytically. In previous
work~\cite{Hirota}, we   considered only the  effect of electron inertia and estimated an 
explosive growth rate by using a new variational method. This method not only gives 
better agreement with simulation results than  earlier work~\cite{Ottaviani},  but also gives a better physical
interpretation because  energy conservation is properly taken into
account.  Here we  generalize our previous study for the Y-shaped layer  and consider 
an X-shaped layer.

To be more specific, we will restrict our consideration to the case of
 $d_e=\rho_s$, for simplicity,  and shorten the scale
 $d_e=\rho_s$ as much as possible in the simulations.  For the first time we perform 
 simulations with $\rho_s=d_e<0.01 L$ and find that the X-shaped
current-vortex layer widens rather locally around the reconnection point
 regardless of the size of computational domain.  We show theoretically
 that this  local X-shaped structure is nonlinearly generated because it is
 optimal for releasing the magnetic energy more efficiently than global structures.

Our variational method~\cite{Hirota} is inspired by the ideal MHD
Lagrangian theory, in which the magnetic energy is considered to be part
of the potential energy of the dynamical system (in analogy with the elastic energy of
rubber bands). If a fluid displacement continually decreases the magnetic
energy, it is likely to grow by gaining the corresponding amount of the
kinetic energy, and the most unstable displacement would decrease the
magnetic energy most effectively.
This argument assumes that the two-fluid effects are essential for changing the
topology of magnetic field lines within the thin boundary layer,
but their impact on the global energy balance is negligible in the limit
of $d_e,\rho_s\rightarrow0$.
By choosing a fluid displacement as a test function that mimics the
local reconnection, we can estimate a growth
rate for the displacement from the kinetic energy.

In Sec.~\ref{sec:model}, we first introduce a reduced
two-fluid model~\cite{Schep} that includes the
effects of electron inertia and electron temperature, and focus on
a collisionless tearing mode that is linearly unstable for a magnetic
shear $B_y(x)\propto\sin(2\pi x/L_x)$ in a doubly
periodic $x$-$y$ plane, where the wavenumber $k_y=2\pi/L_y$ is
related to the aspect ratio $L_y/L_x$ of the domain. In
Sec.~\ref{sec:numerical}, we present our numerical results on the  
nonlinear evolution of this tearing mode. Explosive
growth is observed when $\rho_s\Delta'\gtrsim0.65$ (where $d_e=\rho_s$
and the tearing
index $\Delta'$ is a function of $L_y/L_x$). We will find that the explosive growth rate
is almost independent of $L_y/L_x$ because, in the explosive
phase, the X-shaped current-vortex layer expands locally around the
reconnection point and its characteristic length in the
$y$-direction is shorter than 
 $L_y$.
In Sec.~\ref{sec:theory}, we present our  theoretical model that explains
the generation of such a local X-shaped layer. By replacing $L_y$ with
a shorter length
$\Lambda_y(\le L_y)$, we introduce an effective tearing index $\tilde{\Delta}'$
as a free parameter of the external solution and connect it to
 a novel inner
solution that represents nonlinear evolution of a X-shaped layer.
We will show that a local reconnection $\Lambda_y<L_y$ can transform the
magnetic energy into the kinetic energy more efficiently than the global
one $\Lambda_y=L_y$. 
Using this variational principle  we estimate a
growth rate for this local reconnection model, which is indeed explosive and agrees
with the simulation results. We finally summarize in Sec.~\ref{sec:summary}

\section{Model equations}\label{sec:model}

We begin with  the reduced two-fluid
model given in  Refs.~\cite{Schep,Kuvshinov} with the assumption of cold ions, 
\begin{align}
\frac{\partial\nabla^2\phi}{\partial t}
 +[\phi,\nabla^2\phi]+[\nabla^2\psi,\psi]=0,\label{vorticity}\\
 \frac{\partial\psi}{\partial t}
 +[\phi-\rho_s^2\nabla^2\phi,\psi]
-d_e^2\left(\frac{\partial\nabla^2\psi}{\partial t}+[\phi,\nabla^2\psi]\right)
=0,\label{Ohm}
\end{align}
which governs the two-dimensional velocity field 
$\bm{v}=\bm{e}_z\times\nabla\phi(x,y,t)$
and magnetic field $\bm{B}=\sqrt{\mu_0m_in_0}\nabla\psi(x,y,t)\times\bm{e}_z+B_0\bm{e}_z$, where the guide field $B_0$ and mass density $m_in_0$ are assumed to be constant,  $\mu_0$ is the magnetic
permeability,  and $[f,g]=(\nabla f\times\nabla g)\cdot\bm{e}_z$ is the Poisson bracket.
Here, the effects of electron inertia and electron temperature introduce 
two microscopic scales:  the electron skin depth $d_e=c/\omega_{pe}$ and
 the ion-sound
gyroradius $\rho_s=\sqrt{T_e/m_i}/\omega_{ci}$, respectively
(where $c$ is the speed of light, 
$\omega_{pe}$ is the electron plasma frequency, $\omega_{ci}$ is the ion
cyclotron frequency, $T_e$ is the electron temperature, $m_i$ is the ion mass).

The equations \eqref{vorticity} and \eqref{Ohm} conserve the total energy
(or the Hamiltonian) that is given by the following energy integral,
\begin{align}
 E&=  \frac{1}{2}\int
 d^2x\left(|\nabla\phi|^2+\rho_s^2|\nabla^2\phi|^2+|\nabla\psi|^2+
 d_e^2|\nabla^2\psi|^2\right)\nonumber\\
&=:  E_V+E_T+E_B+E_C,\label{Hamiltonian}
\end{align}
where $E_V= \int|\nabla\phi|^2d^2x/2$ is the ion perpendicular flow
 energy, $E_T= \int\rho_s^2|\nabla^2\phi|^2d^2x/2$ is the electron
 thermal energy, $E_B= \int|\nabla\psi|^2d^2x/2$ is the magnetic energy
 and $E_C= \int d_e^2|\nabla^2\psi|^2d^2x/2$ is the electron parallel flow
 (or current) energy. 
Kuvshinov {\it et al.}~\cite{Kuvshinov2} and Cafaro {\it et
 al.}~\cite{Cafaro} show that \eqref{vorticity} and \eqref{Ohm} can be
 further rewritten as
\begin{align}
\frac{\partial\psi_+}{\partial t}+[\phi_+,\psi_+]=0,\label{conservation1}\\
 \frac{\partial\psi_-}{\partial t}+[\phi_-,\psi_-]=0,\label{conservation2}
\end{align}
in terms of
\begin{align}
 \psi_\pm=\psi-d_e^2\nabla^2\psi\pm\rho_sd_e\nabla^2\phi,\label{psi_pm}\\
 \phi_\pm=\phi-\rho_s^2\nabla^2\phi\pm\rho_sd_e\nabla^2\psi.
\end{align}
It follows that $\psi_+$ and $\psi_-$ are
  frozen-in variables, whereas $\psi$ is not unless $d_e=0$.  Magnetic reconnection is therefore possible when $d_e\ne0$ without
any dissipation mechanism.

As is common with earlier works~\cite{Cafaro,Grasso,Bhattacharjee}, we consider a static equilibrium state,
\begin{align}
\phi^{(0)}(x,y)\equiv0\quad\mbox{and}\quad\psi^{(0)}(x,y)=\psi_0\cos(\alpha
 x), \label{equilibrium}
\end{align}
on a doubly periodic domain $D=[-L_x/2, L_x/2]\times
[-L_y/2,L_y/2]$ (where $\alpha=2\pi/L_x$), which is unstable with
respect to double tearing modes whose reconnection layers are located at
$x=0$ and $x=\pm L_x/2$.
For initial data we assume a sufficiently small perturbation of  a single harmonic, $\phi\propto\sin k_yy$ and $\psi-\psi^{(0)}\propto\cos
k_yy$, where $k_y=2\pi/L_y$. Then, the following parities,
\begin{align}
 \phi(x,y,t)=-\phi(-x,y,t)=-\phi(x,-y,t),\\
 \psi(x,y,t)=\psi(-x,y,t)=\psi(x,-y,t),\label{parity}
\end{align}
are exactly preserved by Eqs.~\eqref{vorticity} and \eqref{Ohm} for all $t$~\cite{Grasso2}.
Therefore, the origin $(x,y)=(0,0)$ (and the four corner points of $D$
       as well) is always a reconnection point.
The solutions to this problem are fully characterized by 
 three parameters;
$d_e/L_x$, $\rho_s/L_x$ and the aspect
ratio $L_y/L_x(=\alpha/k_y)$.

The linear stability of this collisionless tearing mode has been analyzed  in detail by
many authors~\cite{Porcelli,Bhattacharjee,Comisso}. For a given
wavenumber $k_y$ in the $y$-direction, the tearing index
at the reconnection layer $x=0$ is calculated as
\begin{align}
 \Delta'
=2\alpha\sqrt{1-(k_y/\alpha)^2}\tan\left[\frac{\pi}{2}\sqrt{1-(k_y/\alpha)^2}\right],
\end{align}
and the tearing mode is unstable when $\Delta'>0$, namely, $0<k_y/\alpha=L_x/L_y<1$.

For $\rho_s>d_e$, the analytic dispersion relation~\cite{Comisso} predicts that the maximum
growth rate occurs when
\begin{align}
 \Delta'_{\rm max}\sim(2\pi^2)^{1/3}d_e^{-2/3}\rho_s^{-1/3}.\label{max1}
\end{align}
Since $d_e/L_x\ll1$ and $\rho_s/L_x\ll1$ are usually of interest,
this $\Delta'_{\rm max}$ is often very large.
If it belongs to the range
 $L_x\Delta'>100$ (or $k_y/\alpha=L_x/L_y<0.377$) in which $\Delta'$ is well approximated by
\begin{align}
 L_x\Delta'\simeq 16L_y^2/L_x^2=16\alpha^2/k_y^2,
\end{align}
we can estimate the maximum growth rate $\gamma_{\rm max}$ at the wave
number $k_{y,{\rm max}}$ as follows,
\begin{align}
\frac{k_{y,{\rm max}}}{\alpha}\simeq
 \sqrt{\frac{16}{L_x\Delta'_{\rm max}}}
\sim 2.43\sqrt{\frac{d_e^{2/3}\rho_s^{1/3}}{L_x}},\label{max2}\\
 \gamma_{\rm max}\sim (2/\pi)^{1/3}\frac{k_{y,{\rm max}}d_e^{1/3}\rho_s^{2/3}}{\tau_H}
\sim\frac{13.1}{\tau_H}\frac{d_e^{2/3}\rho_s^{5/6}}{L_x^{3/2}},\label{max3}
\end{align}
where $\tau_H^{-1}=\psi_0\alpha^2$.

Given this background material we now turn to our numerical simulations. 

\section{Numerical results}\label{sec:numerical}

Equations \eqref{vorticity} and \eqref{Ohm} are solved numerically
for  various parameters using the spectral
method in both the $x$ and $y$ directions and 4th-order Runge-Kutta
 for time evolution. The nonlinear acceleration phase is 
observed when the magnetic island width becomes larger than the
reconnection layer width that  is of order $d_e^{2/3}\rho_s^{1/3}$ for
the case $\rho_s\ge d_e$. To observe this phase for a longer period,
 we have performed all simulations with $d_e=\rho_s$ and narrowed  
 the layer width ($\sim d_e=\rho_s$) as much as possible.
The most demanding case $d_e=\rho_s=0.005L_x$ requires $8192\times8192$
grid points in   wavenumber space.

\begin{figure}
 \includegraphics[width=8cm]{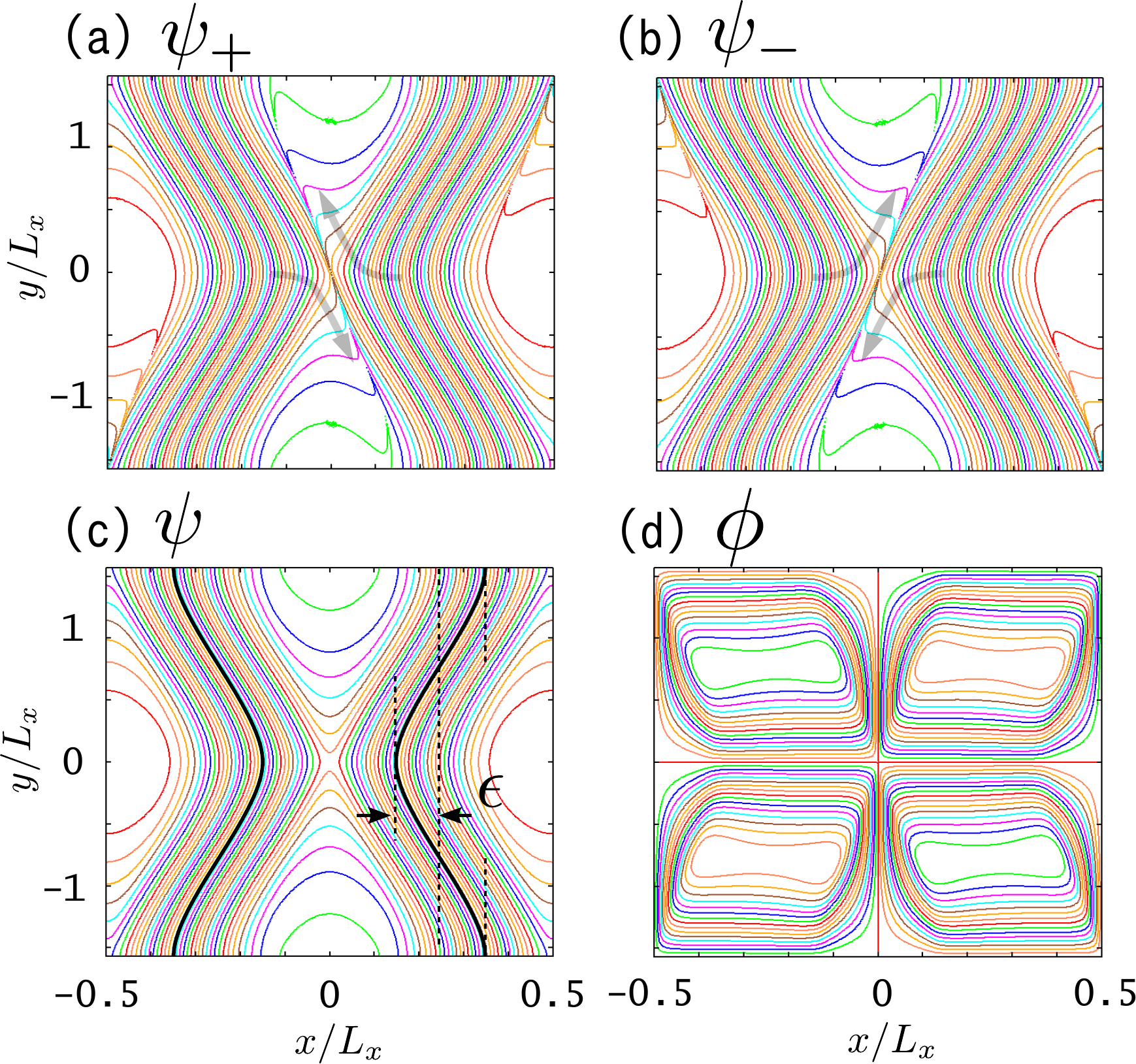}
 \caption{Contours of $\psi_\pm$, $\psi$ and $\phi$ when $\epsilon=5\rho_s$, where $\rho_s=d_e=0.02L_x$, $L_y/L_x=\pi$.}
 \label{contours1}
\end{figure}

\begin{figure}
 \includegraphics[width=8cm]{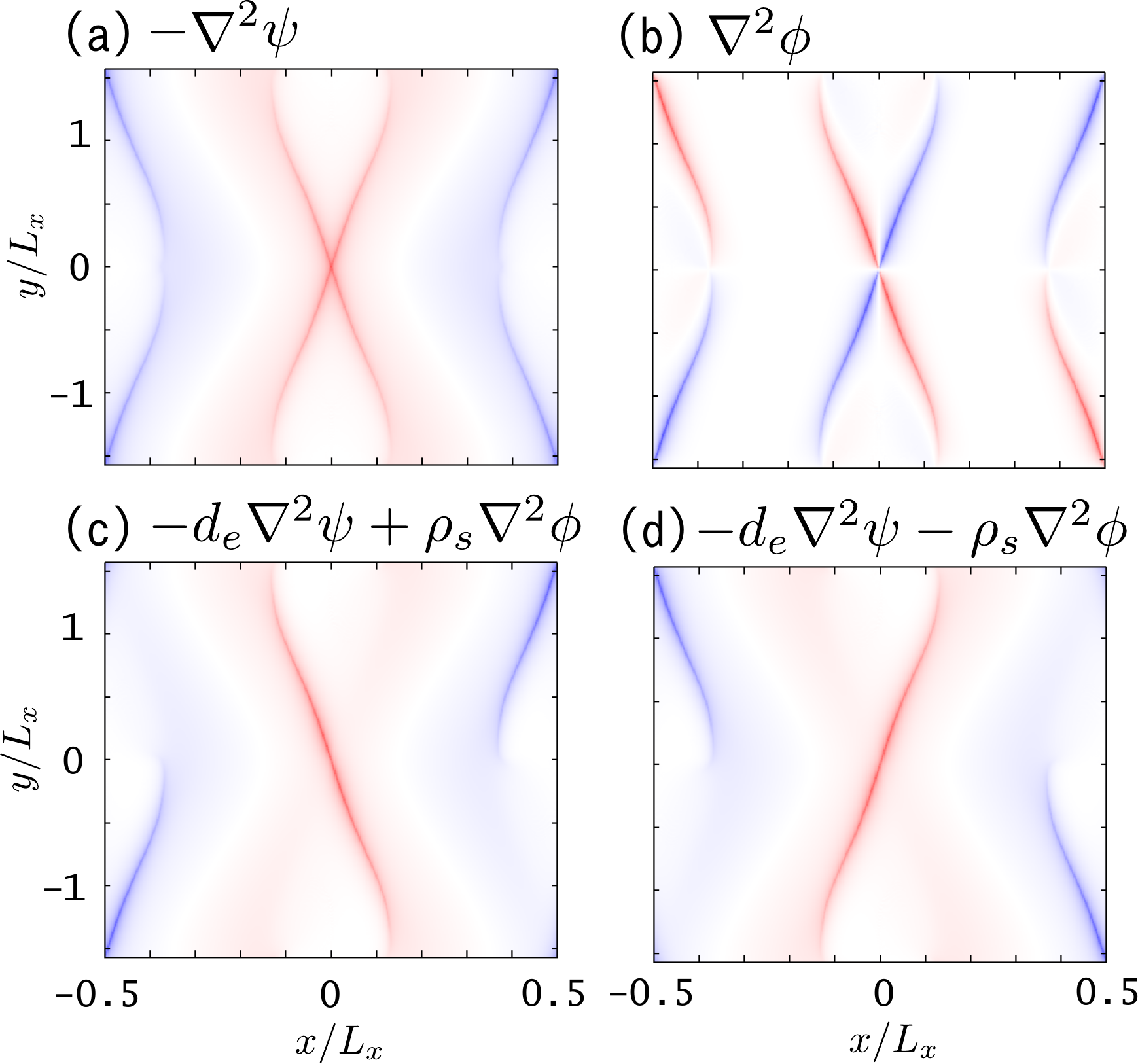}
 \caption{Intensity distributions (red: positive, blue: negative) of the current $-\nabla^2\psi$,
 vorticity $\nabla^2\phi$ and $-d_e\nabla^2\psi\pm\rho_s\nabla^2\phi$ when $\epsilon=5\rho_s$, 
 where $\rho_s=d_e=0.02L_x$, $L_y/L_x=\pi$.}
 \label{contours2}
\end{figure}

The nonlinear evolution of \eqref{vorticity} and \eqref{Ohm}
was  studied in  earlier works~\cite{Cafaro,Grasso,Grasso2,Bhattacharjee},  and we reproduce the 
main features,  as shown in Figs.~\ref{contours1} and \ref{contours2}.
Since $\psi_+$ and $\psi_-$ are   frozen-in variables, their contours preserve
topology as  seen in  Figs.~\ref{contours1}(a) and (b) (where the arrows
depict the fluid motions generated by $\phi_\pm$).
Then, spiky peeks of $\psi_+$ and $\psi_-$ are formed and their ridge lines look
like the shapes of ``$\backslash$'' and ``/'', respectively, around the origin.
In light of the definition \eqref{psi_pm}, the current and vorticity
distributions can be directly calculated by using
$\psi_\pm$ and shown in Figs.~\ref{contours2}(a) and (b), which exhibit a ``X''-shaped
current-vortex layer~\cite{Aydemir,Cafaro} whose width is of
order $d_e=\rho_s$. We also note from Figs.~\ref{contours2}(c) and (d) that
a relation $d_e|\nabla^2\psi|\simeq\rho_s|\nabla^2\phi|$ holds inside
the layer (except at the reconnection points).

\begin{figure}
 \includegraphics[width=8cm]{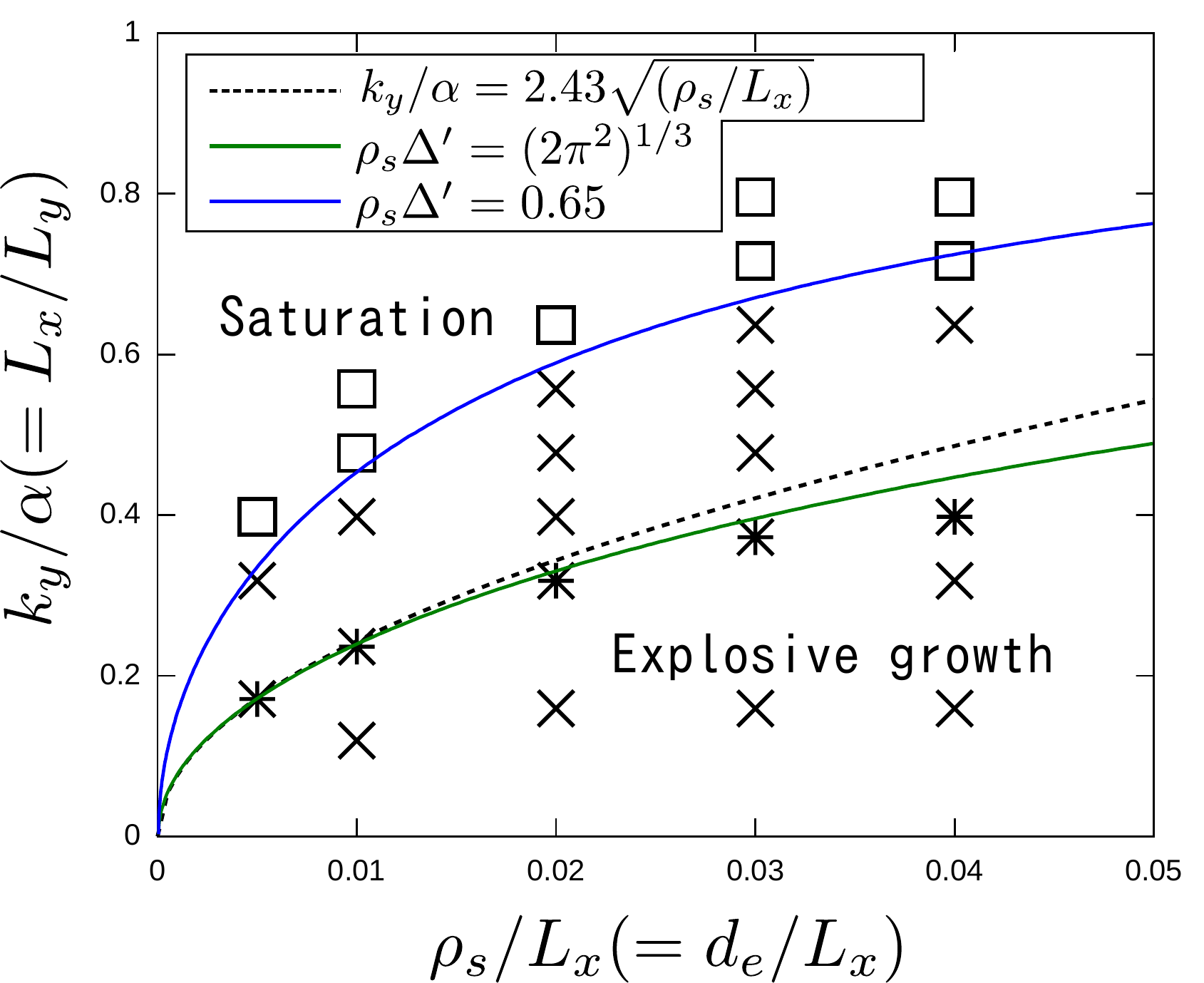}
 \caption{Parameters $k_y$ and $\rho_s(=d_e)$ that result in explosive
 growth ($\times$ and $*$) and saturation ($\square$).  The maximum linear
 growth rate occurs at  points  $*$ for each $\rho_s$.}
 \label{diagram}
\end{figure}

As indicated in Fig.~\ref{contours1}(c), we denote the maximum
displacement of the field lines at $x=\pm L_x/4$ by $\epsilon$ and numerically measure
it from the displacement of the contour $\psi=0$. 
We have run simulations with various combinations of $L_x/L_y$ and
$\rho_s/L_x(=d_e/L_x)$, and investigated whether $\epsilon$ grows explosively
or not. Our results are summarized in Fig.~\ref{diagram}, where it should be
recalled that the tearing mode is linearly unstable at all points in
this figure since $0<L_x/L_y<1$. The linear growth rate achieves its numerical  maximum  
at points indicated by the asterisk ($*$)   for each fixed $\rho_s$, which agrees
with the theoretical prediction \eqref{max1} [and also
\eqref{max2} for $k_y/\alpha<0.377$]. 
The square symbol ($\square$)  indicates points where  exponential growth $\epsilon\propto
e^{\gamma t}$ (with the linear growth rate $\gamma$) stalls before $\epsilon$ reaches $\rho_s$.
On the other hand, at the crosses  ($\times$)
and asterisks ($*$) the exponential growth  is
accelerated when $\epsilon$ gets larger than $\rho_s$.
These two regimes seem to be divided by a curve $\rho_s\Delta'\sim0.65$.

\begin{figure}
 \includegraphics[width=10cm]{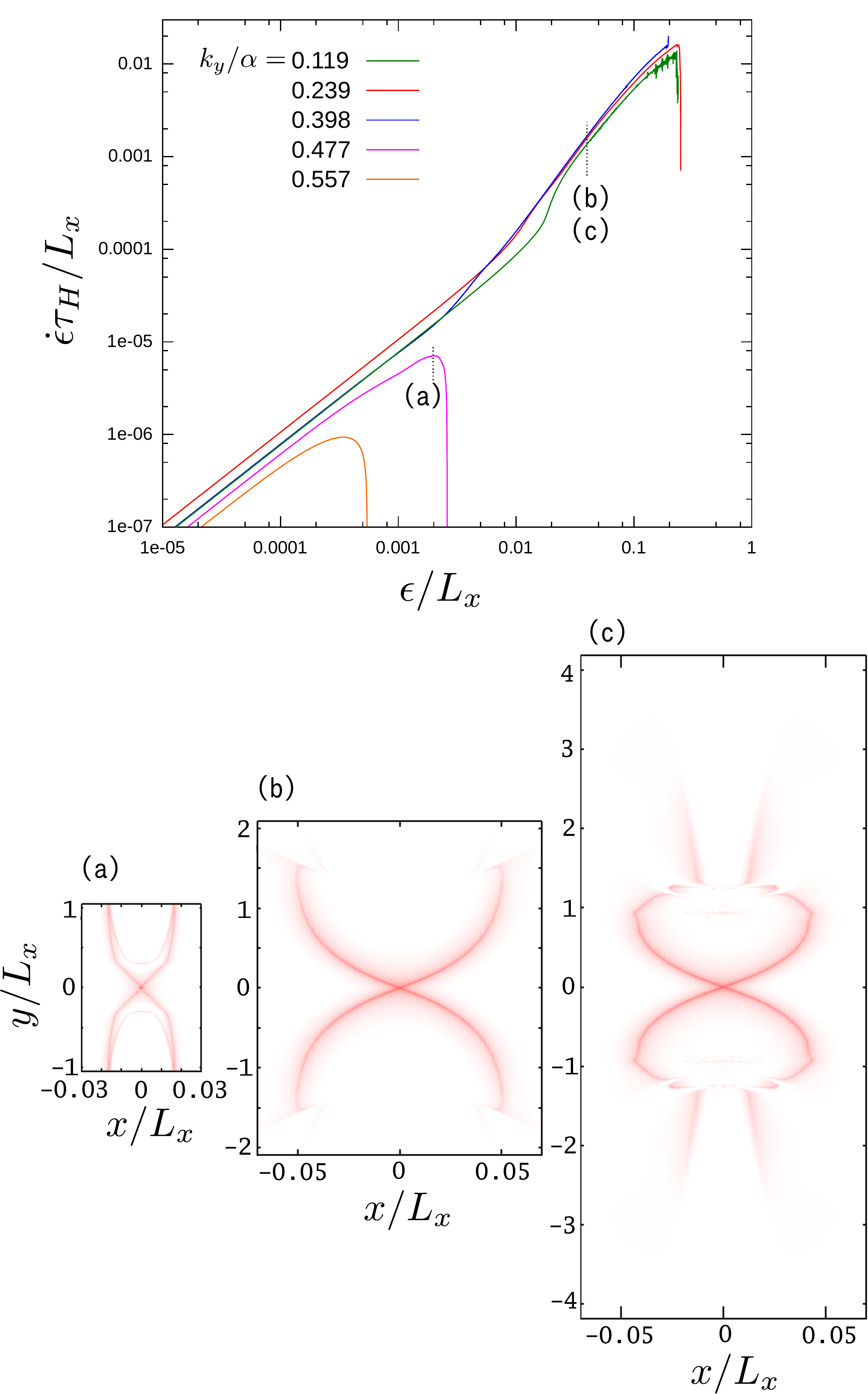}
 \caption{Logarithmic plots of the displacement $\epsilon$ versus its
 time-derivative $\dot{\epsilon}$, 
where $d_e=\rho_s=0.01$, and the current distributions (a) at
 $\epsilon=0.002L_x$ for $k_y/\alpha=0.477$, (b)
at $\epsilon=0.04L_x$ for $k_y/\alpha=0.239$ and 
(c) at $\epsilon=0.04L_x$ for $k_y/\alpha=0.119$.}
\label{growth_aspect}
\end{figure}

The above mentioned tendencies are demonstrated in
Fig.~\ref{growth_aspect}, which is
 a logarithmic plot of $\dot{\epsilon}$ versus $\epsilon$ for
the case  $\rho_s/L_x=0.01$.
For $k_y/\alpha=0.477$ and $0.557$, which belong to the saturation regime $\rho_s\Delta'\lesssim0.65$, the current-vortex layer spirals around the O-point as
 shown in Fig.~\ref{growth_aspect}(a) and the growth of $\epsilon$ decelerates.
Although this occurs in an early nonlinear phase $\epsilon<\rho_s$ in our results,
we note that the saturation mechanism is similar to the one found by Grasso {\it
 et al.}~\cite{Grasso2}, namely,
the phase mixing of the Lagrangian (or frozen-in) invariants $\psi_\pm$ leads to a new ``macroscopic'' stationary state.
For the cases of $k_y/\alpha=0.119$, 0.239, and 0.398,  which belong to $\rho_s\Delta'\gtrsim0.65$, we observe a transition from
the exponential growth $\dot{\epsilon}\propto\epsilon$ to an explosive
growth $\dot{\epsilon}\propto\epsilon^n$ ($n>1$) around
$\epsilon\sim\rho_s$, and the latter continues until the reconnection
completes at $\epsilon=L_x/4$.
We further note that, for the small $k_y/\alpha=0.119$, a local X-shaped layer is generated spontaneously around
the reconnection point and it expands
faster than the global one that originates from the linear eigenmode
[see Fig.~\ref{contours2}(c)].
By comparing the case $k_y=0.119$ with $k_y=0.239$ at the same amplitude $\epsilon=0.04L_x$, we find that this local X-shaped structure
around the origin in
Fig.~\ref{contours2}(c) is identical to that in Fig.~\ref{contours2}(b).
Therefore, the explosive reconnection seems to be
attributed to the fast expansion of the local X-shape with a certain optimal size that is
independent of the domain length $L_y=2\pi/k_y$.
In fact, the nonlinear growth rates for $k_y=0.119$, 0.239, and $0.398$ are
eventually comparable for $\epsilon\gtrsim0.02$
because the released magnetic energies are almost the same
regardless of $k_y$.

\begin{figure}
 \includegraphics[width=7cm]{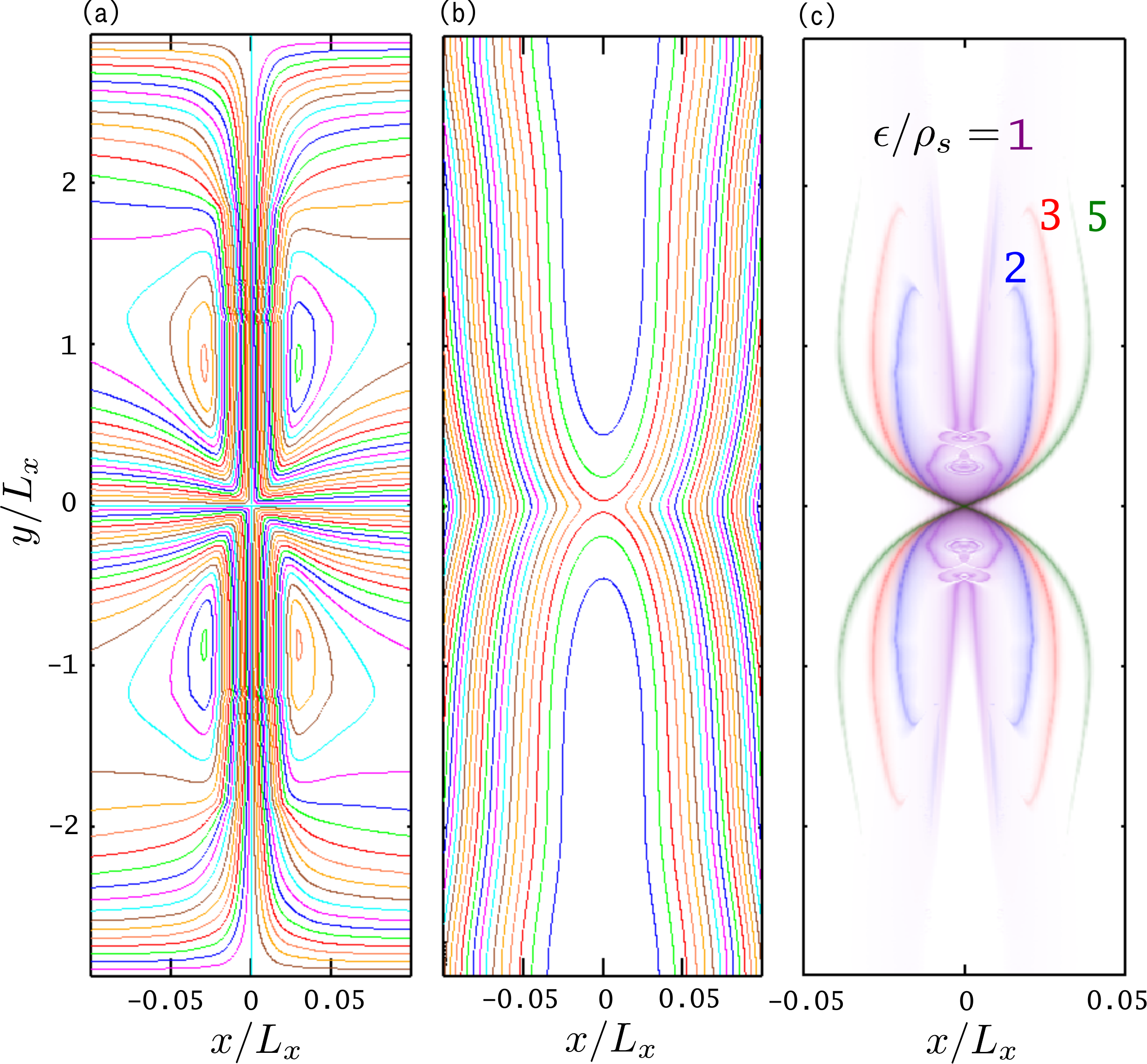}
 \caption{Contours of (a) $\phi$ and (b) $\psi$ at
 $\epsilon=0.015L_x=3\rho_s$ and (c) current distributions at $\epsilon/\rho_s=1,2,3,5$, where $d_e=\rho_s=0.005L_x$, $k_y/\alpha=k_{y,{\rm max}}/\alpha=0.171$.}
 \label{contours3}
\end{figure}

\begin{figure}
 \includegraphics[width=7cm]{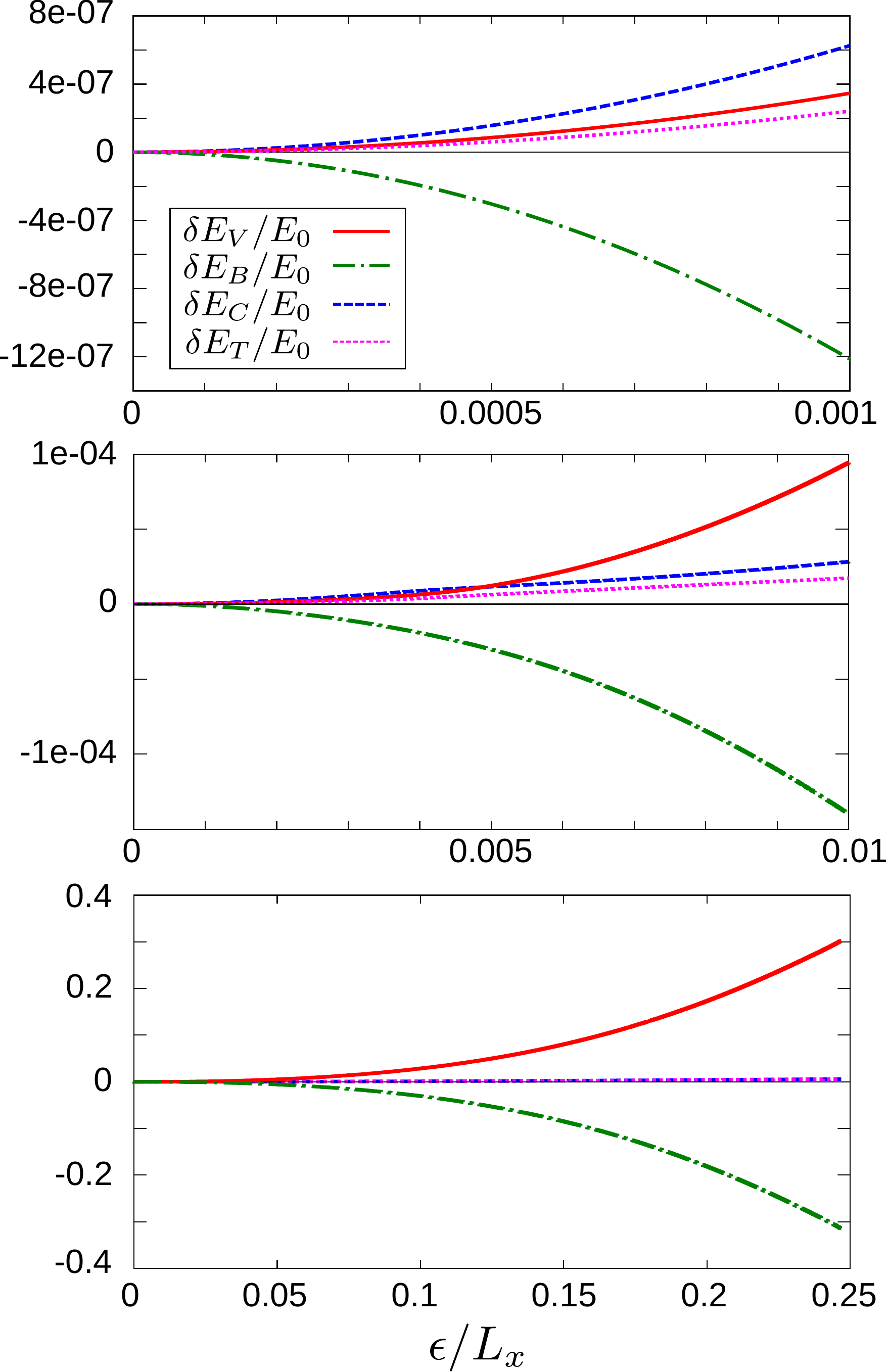} 
 \caption{Changes of energies $E_V,E_B,E_C$ and $E_T$ versus $\epsilon$, where $d_e=\rho_s=0.005L_x$, $k_y/\alpha=k_{y,{\rm max}}/\alpha=0.171$.}
 \label{energy}
\end{figure}

We remark that the length of the local X-shape is not simply related to
the wavelength $2\pi/k_{y,{\rm max}}$ of the most linearly unstable
mode. For $d_e=\rho_s=0.005L_x$
and $k_y=k_{y,{\rm max}}=0.171\alpha$,
Figs.~\ref{contours3}(a) and (b) give a closer look at the contours of
$\phi$ and $\psi$, respectively, at
$\epsilon=3\rho_s$, and
 Fig.~\ref{contours3}(c) shows the shapes of the current layer 
observed at $\epsilon/\rho_s=1,2,3$, and 5. As can be seen from
Fig.~\ref{contours3}(c), a local X-shape appears 
and expands quickly in the nonlinear phase $\epsilon>\rho_s$ even though
this reconnection is triggered by the most linearly unstable tearing mode
$k_y=k_{y,{\rm max}}$. Under the same conditions,
the evolution of the energies $E_{V,T,B,C}$ defined by
\eqref{Hamiltonian} is shown in
Fig.~\ref{energy} (where the total energy conservation $E=E_0=$ const. is
satisfied numerically with  sufficient accuracy).
In the linear phase $\epsilon\ll\rho_s=0.005L_x$, the magnetic energy $E_B$ is
transformed into $E_V,E_T$ and $E_C$ at different but comparable
rates. However, in the nonlinear phase $\epsilon>\rho_s$, we note that the
magnetic energy is exclusively transformed into the ion flow energy
$E_V$ and
the energy balance $\delta E_V+\delta E_B\simeq0$ is satisfied approximately. Since $\delta E_T$ and $\delta E_C$ are negligible, we infer
that the nonlinear dynamics is dominantly governed by the ideal MHD
equation ($d_e,\rho_s\rightarrow0$).
This fact motivates us to regard $K:=E_V$ and
$W:=E_B$ as the
kinetic and potential energies, respectively, according to the MHD Lagrangian theory.

\section{Theoretical model}\label{sec:theory}

In this section, we develop a theoretical model to explain the
explosive growth of $\epsilon$ in the nonlinear phase
$d_e=\rho_s\ll\epsilon\ll L_x/4$.
The current-vortex layers are obviously the boundary layers caused by
the two-fluid effects and their width should be of order $d_e=\rho_s$.
The ideal MHD equations, \eqref{vorticity} and \eqref{Ohm} with
$d_e=\rho_s=0$, would be satisfied approximately outside the
boundary layers.
Moreover, we also assume that $\phi$ and $\psi$ are continuous across the
boundary layers, because $E_C= \int d_e^2|\nabla^2\psi|^2d^2x/2$ and
$E_T= \int \rho_s^2|\nabla^2\phi|^2d^2x/2$ are negligible in the
energy conservation (Fig.~\ref{energy}) in the limit of
$d_e,\rho_s\rightarrow0$.
 Note that $\nabla\psi$ and $\nabla\phi$ may be
discontinuous across the layer because it is a current-vortex layer.

\begin{figure}
 \includegraphics[width=7cm]{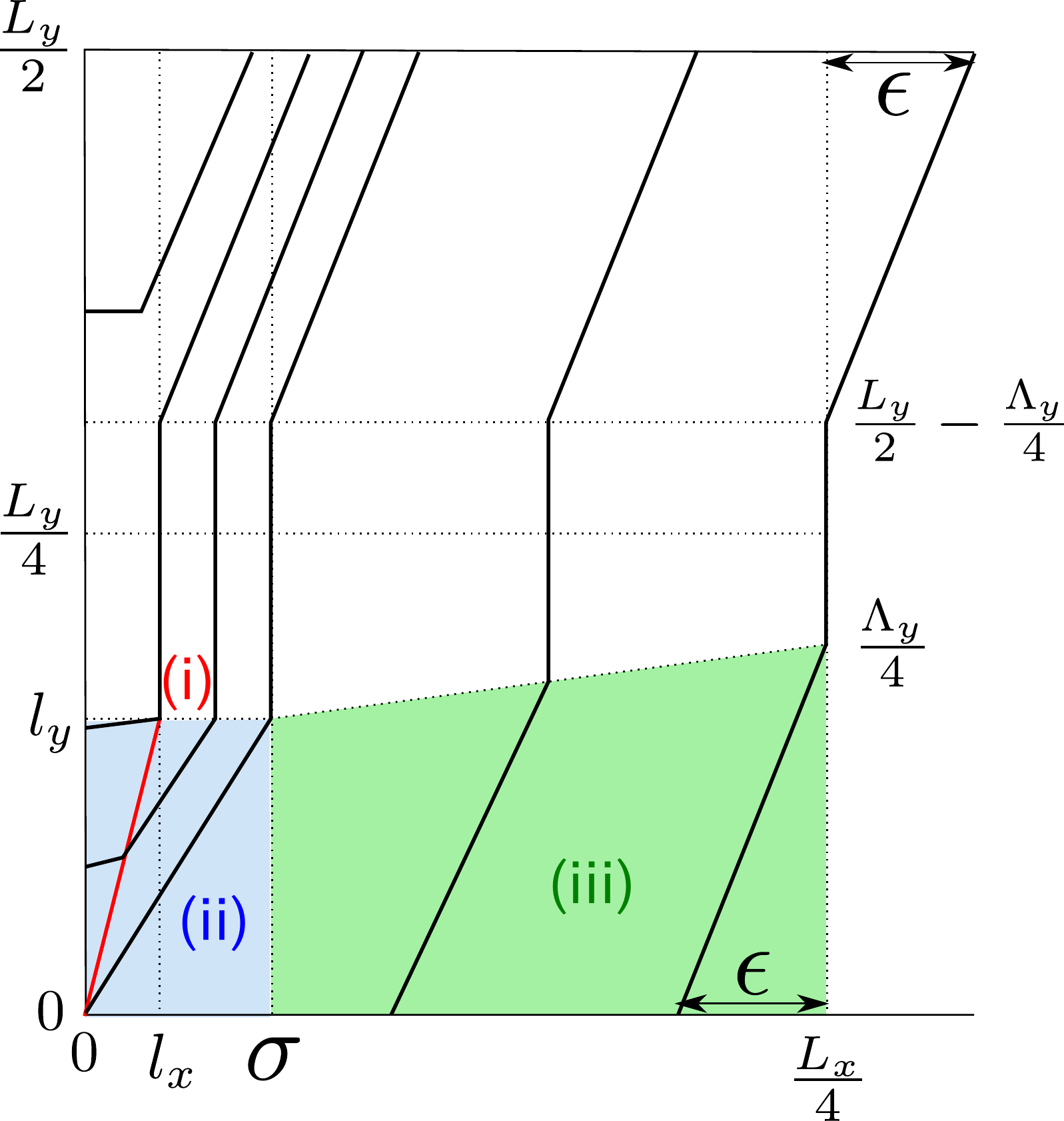}
 \caption{Local reconnection model.}
 \label{Local_reconnection}
\end{figure}

Based on these assumptions, we consider a family of virtual
displacements that generates a local X-shaped current-vortex layer, and
then seek the displacement that decreases the magnetic (or potential) energy
most effectively.
Our reconnection model is illustrated in
Fig.~\ref{Local_reconnection}, where 
we show only the first quadrant around the origin owing to
the parity~\eqref{parity}.
In Fig.~\ref{Local_reconnection}, the magnetic field lines are assumed to be
piecewise-linear and the red line denotes the boundary layer (i.e.,
the upper right part of the X-shape). 

We will mainly focus on the three regions: (i) boundary layer (ii) inner
solution with a X-shaped layer (iii) external solution.  These regions
are characterized by four parameters $\Lambda_y,\sigma,l_x,l_y$ as
follows.  First, the position of the boundary layer is specified by
$(l_x,l_y)$. Second, the displacement of the field line that is about to
reconnect at the origin is denoted by $\sigma$, which can be also
regarded as the half width of the local island.  Finally, to allow for  local reconnection, we introduce
the ``wavelength'' $\Lambda_y$ of the displacement at $x=L_x/4$, which
may be smaller than the wavelength $L_y(=2\pi/k_y)$ of the linear
tearing mode.
We will assume the following orderings among these parameters: 
\begin{align}
 d_e=\rho_s\ll\epsilon\le\sigma\ll L_x/4,\quad
 l_x<\sigma\ll l_y\le\Lambda_y/4\le L_y/4.\label{orderings}
\end{align}

\subsection{Matching conditions across the boundary layer}

\begin{figure}
\includegraphics[width=6cm]{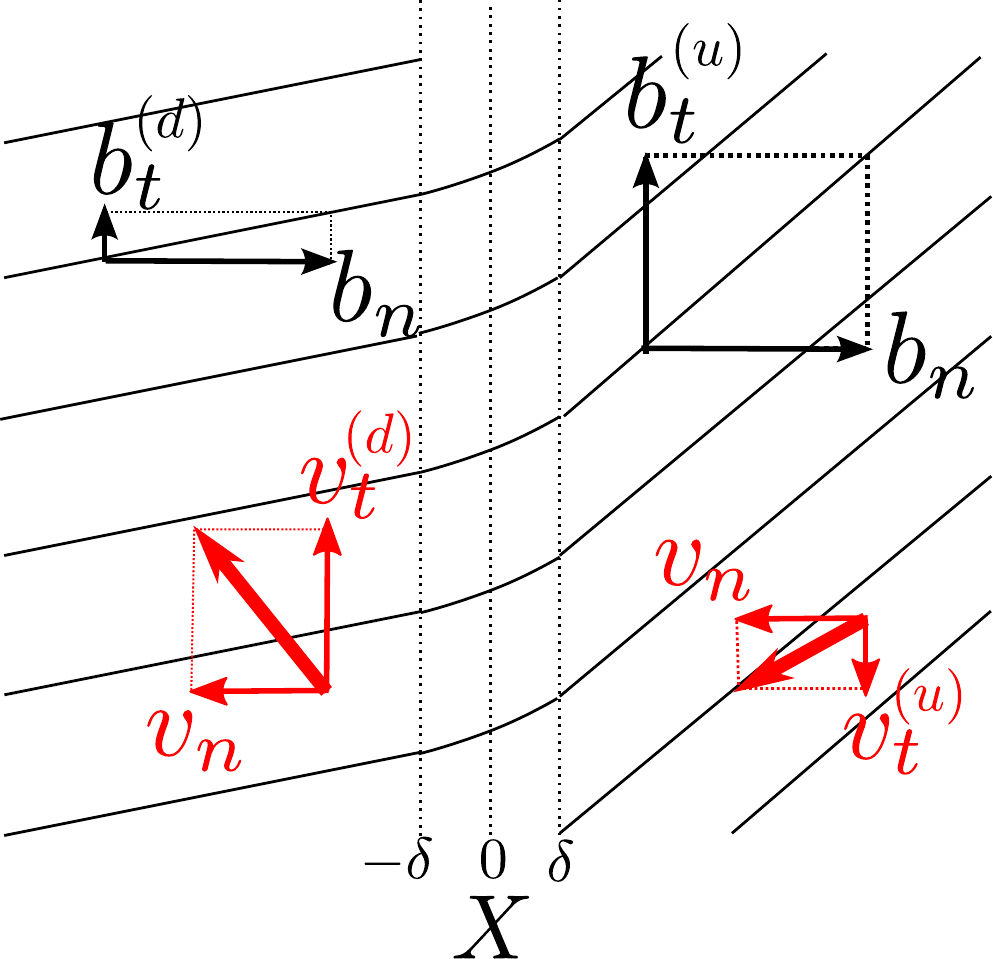}
\caption{Neighborhood of the current-vortex layer.}\label{fig:boundary} 
\end{figure}

First, we focus on a neighborhood of (i) the boundary layer and
introduce a local coordinate system $(X,Y)$ in the frame moving with the
boundary layer, so that the $X$ and $Y$ directions are respectively normal
and tangent to the layer (see Fig.~\ref{fig:boundary}). Let the inner region of the boundary layer be
$-\delta<X<\delta$, where $\delta\sim d_e=\rho_s$.
In this coordinate system, the velocity $\bm{v}$ and the Alfv\'en velocity
$\bm{b}=\bm{B}/\sqrt{\mu_0\rho_0}$ are assumed to be uniform outside the layer.
Using the continuities of $\psi$ and $\phi$ across the layer,
we assign linear functions,
       \begin{align}
	\begin{cases}
	 \psi=\psi_c-b_t^{(d)}X+b_nY,\\	 
	 \phi=\phi_c+v_t^{(d)}X-v_nY,	 
	\end{cases}\label{downstream}
        \end{align}
on the down-stream side ($X<-\delta$) and
\begin{align}
\begin{cases}
 \psi=\psi_c-b_t^{(u)}X+b_nY,\\
 \phi=\phi_c+v_t^{(u)}X-v_nY,
\end{cases}\label{upperstream}
\end{align}
on the up-stream side ($X>\delta$), where
 all coefficients depend only on time. The discontinuities of the tangential components,
$b_t^{(d)}\ne b_t^{(u)}$ and $v_t^{(d)}\ne v_t^{(u)}$, indicate 
the presence of a current-vortex layer within $[-\delta,\delta]$.

Since we have assumed that $\partial_t\psi+[\phi,\psi]=0$ holds outside the
layer, $[\phi,\psi]$ must be also continuous, namely,
\begin{align}
v_t^{(d)}b_n-v_nb_t^{(d)}=v_t^{(u)}b_n-v_nb_t^{(u)},\label{IMHD0}
\end{align}
is one of the matching conditions between \eqref{downstream} and \eqref{upperstream}.

Moreover, the conservation laws of $\psi_\pm$ require that
$\psi_-$ has a spiky peek within the boundary layer
whereas $\psi_+$ does not [see Figs.~\ref{contours1}(a) and (b)]. 
Since $\psi_+\simeq\psi$ holds, we find a relation,
\begin{align}
d_e\nabla^2\psi\simeq\rho_s\nabla^2\phi\quad\mbox{on}\quad [-\delta,\delta],
\end{align}
which specifies the ratio between a current peak ($-\nabla^2\psi$) and
a vorticity peak ($\nabla^2\phi$) inside the layer.
We have already noticed this relation in Fig.~\ref{contours2}.
By integrating these current and vorticity distributions over
$[-\delta,\delta]$, we obtain another matching condition,
\begin{align}
 -\rho_s(v_t^{(u)}-v_t^{(d)})=d_e(b_t^{(u)}-b_t^{(d)}). \label{EMHD}
\end{align}

When the boundary layer is moving at a speed $V_n$ in the $X$
direction, the condition \eqref{IMHD0} is
 transformed to
\begin{align}
v_t^{(d)}b_n-(v_n-V_n)b_t^{(d)}=v_t^{(u)}b_n-(v_n-V_n)b_t^{(u)}, \label{IMHD}
\end{align}
in the rest frame, and the condition \eqref{EMHD} is unchanged.
Since these conditons also yield
\begin{align}
 \rho_s(v_n-V_n)=-d_eb_n,\label{IMHD2}
\end{align}
 we need to impose at least two matching conditions among \eqref{EMHD},
 \eqref{IMHD} and \eqref{IMHD2}.

\subsection{Modeling of the X-shaped boundary layer}

\begin{figure}
\includegraphics[width=10cm]{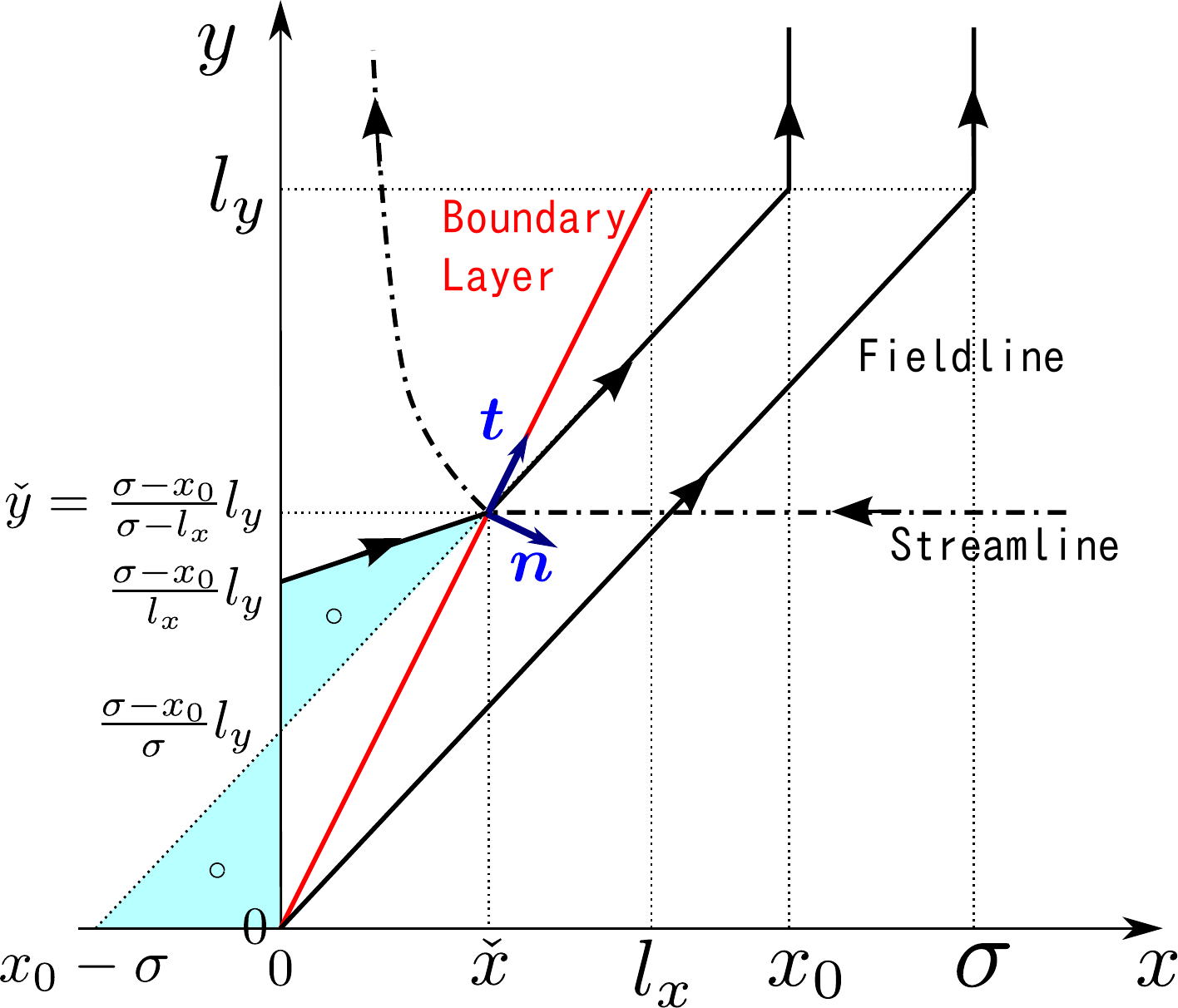} 
\caption{First quadrant of the X-shaped boundary layer.}\label{fig:X-shape}
\end{figure}

Next, we consider (ii) the inner solution that contains the X-shaped
boundary layer. The detailed sketch
of this region is given in Fig.~\ref{fig:X-shape}, 
where the displacement map
$(x_0,y_0)\mapsto(x,y)$ on the up-stream side (i.e., the right side of the boundary layer) is
simply modeled by
\begin{align}
 x=x_0+\frac{\sigma}{l_y}(y_0-l_y)
\quad\mbox{ and }\quad
y=y_0.
\end{align}

This displacement map fully determines $\psi$ and $\phi$ on the
up-stream side as follows.
Since $\sigma\ll L_x/4$ is assumed in \eqref{orderings}, we expand the equilibrium flux function,
\begin{align}
 \psi^{(0)}(x,y)=\psi_0\cos(\alpha x)=\psi_0\left(1-\frac{\alpha^2}{2}x^2\right)+O(x^3),
\end{align}
and neglect $O(x^3)$ in this region $[0,\sigma]$.
Except on the boundary layer, the magnetic flux $\psi$ is frozen into the displacement and hence becomes
\begin{align}
 \psi^{(u)}(x,y,t)=\psi^{(0)}(x_0,y_0)=&\psi_0-\psi_0\frac{\alpha^2}{2}\left[x-\frac{\sigma}{l_y}(y-l_y)\right]^2,
\end{align}
on the up-stream side. 
By regarding the parameters $\sigma(t),l_x(t),l_y(t)$ as functions of
time, the time derivative of the displacement map gives the stream function,
\begin{align}
 \phi^{(u)}(x,y,t)=-\dot{\mathsf{s}}_y\frac{y^2}{2}+\dot{\sigma}y,
\end{align}
where $\mathsf{s}_y=\sigma/l_y$,
and the parity $\phi(x,0,t)=0$ has been used as the boundary condition.

Now, let us consider a magnetic field line that is labeled by its
initial position $x=x_0$ (where $l_x<x_0<\sigma$). When this field line is
displaced by the map, it intersects with the boundary layer at
\begin{align}
 (\check{x},\check{y})=\left(\frac{\sigma-x_0}{\sigma-l_x}l_x,\frac{\sigma-x_0}{\sigma-l_x}l_y\right).
\end{align}

The tangent and normal unit vectors to the boundary layer are respectively
given by $ \bm{t}=(l_x,l_y)/|\bm{l}|$ and $\bm{n}=(l_y,-l_x)/|\bm{l}|$,
where $|\bm{l}|=\sqrt{l_x^2+l_y^2}$.
Therefore, the normal and tangent components at $(\check{x},\check{y})$
are calculated as follows;
\begin{align}
 b_n^{(u)}=&\tau_H^{-1}(\sigma-l_x)\frac{x_0}{|\bm{l}|}\quad(>0),\\
 b_t^{(u)}=&\tau_H^{-1}\left(\frac{l_x}{l_y}\sigma+l_y\right)\frac{x_0}{|\bm{l}|}\quad(>0),\\
 v_n^{(u)}=&\frac{l_y}{|\bm{l}|}(\dot{\mathsf{s}}_y\check{y}-\dot{\sigma})\quad(<0),\\
 v_t^{(u)}=&\frac{l_x}{|\bm{l}|}(\dot{\mathsf{s}}_y\check{y}-\dot{\sigma})\quad(<0),
\end{align}
where $\tau_H^{-1}:=\psi_0\alpha^2$.

Next, we consider the down-stream side, on which the magnetic field
lines are again approximated
by straight lines as shown in Fig.~\ref{fig:X-shape}. Since the
displacement map is area-preserving, the same field line that passes
through $(\check{x},\check{y})$ is found to be
\begin{align}
 y=\frac{l_y}{l_x}\left(\frac{2l_x-\sigma}{l_x}x+\sigma-x_0\right),\label{fieldline}
\end{align}
by equating the areas of the two blue triangles in Fig.~\ref{fig:X-shape}.
Using the fact that the value of $\psi$ is again
$\psi^{(0)}(x_0,y_0)$ on this field line, a straightforward calculation results in
\begin{align}
 \psi^{(d)}(x,y,t)=&\tau_H^{-1}\frac{l_x}{l_y}\left(\sigma y-\frac{l_x}{l_y}\frac{y^2}{2}
+\frac{2l_x-\sigma}{l_x}xy\right)\nonumber\\
&+\tau_H^{-1}\frac{2l_x-\sigma}{l_x}\left(-\sigma x-\frac{2l_x-\sigma}{l_x}\frac{x^2}{2}\right)+\psi_0-\tau_H^{-1}\frac{\sigma^2}{2}.
\end{align}
The field line \eqref{fieldline} also moves in time
because of the time dependence of $l_x,l_y$ and $\sigma$. By imposing 
the boundary condition $\phi(0,y,t)=0$ on the $y$ axis, the associated
incompressible flow can be determined uniquely as
\begin{align}
 \phi^{(d)}(x,y)&=-\frac{\dot{\mathsf{s}}_y-2\dot{\beta}}{\beta}\frac{x^2}{2}
-\frac{\dot{\beta}}{\beta}xy
+\frac{\dot{\sigma}}{\beta}x,\label{outflow}
\end{align}
where $\beta=l_x/l_y$.
We thus obtain, at $(\check{x},\check{y})$,
\begin{align}
 b_t^{(d)}=&\tau_H^{-1}\left(\frac{l_x^2}{l_y}
+\frac{2l_x-\sigma}{l_x}l_y\right)\frac{x_0}{|\bm{l}|},\\
 v_t^{(d)}=&\frac{l_x}{|\bm{l}|}\left[
\left(\frac{|\bm{l}|^2}{l_x^2}\dot{\beta}-\frac{l_y^2}{l_x^2}\dot{\mathsf{s}}_y\right)\check{y}
+\frac{l_y^2}{l_x^2}\dot{\sigma}
\right],
\end{align}
and confirm that $b_n^{(d)}=b_n^{(u)}$ and $v_n^{(d)}=v_n^{(u)}$ are
indeed satisfied.

The   speed $V_n$ for movement of the boundary layer at $(\check{x},\check{y})$
is calculated by using 
 the angle $\theta$ between the boundary layer and the $y$
axis ($\tan\theta=l_x/l_y$), 
\begin{align}
 V_n=\sqrt{\check{x}^2+\check{y}^2}\dot{\theta}
 =\frac{l_y}{|\bm{l}|}\dot{\beta}\check{y}.
\end{align}

Now, we are ready to impose the matching conditions on these up- and
down-stream solutions.
It is interesting to note that the matching condition \eqref{IMHD} is already
satisfied because we have taken the continuities of $\psi$ and $\phi$
into account in the above construction.
The matching condition \eqref{EMHD} at $(\check{x},\check{y})$ gives
\begin{align}
\dot{\zeta}\frac{\sigma-x_0}{\zeta}-\dot{\sigma}=-\tau_H^{-1}x_0\zeta,
\end{align}
where  $\zeta:=\mathsf{s}_y-\beta=(\sigma-l_x)/l_y$.
This condition must be satisfied for all points
$(\check{x},\check{y})$ on the boundary layer (that is, for all
$x_0\in[l_x,\sigma]$), which requires both
$\dot{\zeta}\sigma=\dot{\sigma}\zeta$ and
$\dot{\zeta}=\tau_H^{-1}\zeta^2$ to be satisfied.
The former gives a constant of motion,
\begin{align}
 \frac{\sigma}{\zeta}=\frac{\sigma}{\sigma-l_x}l_y=l_{y0}=\mbox{const.},
\end{align}
and the latter gives an evolution equation,
\begin{align}
\dot{\sigma}=\frac{1}{l_{y0}\tau_H}\sigma^2.\label{sigma}
\end{align}
Although the constant $l_{y0}(>0)$ is still unknown unless $l_x$ and $l_y$
are specified, the displacement $\sigma$ turns out to grow explosively
due to the presence of the X-shaped boundary layer.
These parameters $l_x$, $l_y$ and $l_{y0}$ will be determined later when
this inner solution is matched with the external solution
and the global energy balance is taken into account.

\subsection{External solution}

Now consider  (iii),  the external solution of  Fig.~\ref{Local_reconnection}.
Even though we discuss the nonlinear phase, the displacement
$\sigma$ (or the island half-width) must   be small
$\sigma\ll L_x/4$  as  well as  the growth rate
$\dot{\sigma}/\sigma\ll\tau_H^{-1}$ in comparison with the equilibrium
space-time scale. Therefore, we expect the external
solution to be similar to the well-known eigenfunction of the linear
tearing mode.
This treatment for the external solution is commonly used in
Rutherford's theory~\cite{Rutherford,White}, while we introduce 
 the arbitrary wavelength $\Lambda_y(<L_y)$ of the linear tearing
mode in this work.
Namely, the displacement in the $x$ direction is given by $\xi(x,y)=\epsilon\hat{\xi}(x)\cos(2\pi y/\Lambda_y)$
for $y\in[0,\Lambda_y/4]$ where  the eigenfunction,
\begin{align}
 \hat{\xi}(x)=-\frac{\cos\left[\sqrt{1-(L_x/\Lambda_y)^2}\left(\alpha|x|-\frac{\pi}{2}\right)\right]}{\sin\alpha
 x},
\end{align}
 is normalized so as to satisfy
$\hat{\xi}(L_x/4)=-1$ and
$\hat{\xi}(-L_x/4)=1$.
Taylor expansion of $\hat{\xi}$ around $x=0$ on the positive side
($x>0$) gives 
\begin{align}
 \hat{\xi}(x)=&-\left(\frac{2}{\tilde{\Delta}'x}+1\right)\sqrt{1-(L_x/\Lambda_y)^2}\cos\left[\sqrt{1-(L_x/\Lambda_y)^2}\frac{\pi}{2}\right]+O(x),
\end{align}
where the tearing index $\tilde{\Delta}'$ for the wavelength $\Lambda_y$ is
\begin{align}
 \tilde{\Delta}'
=2\alpha\sqrt{1-(L_x/\Lambda_y)^2}\tan\left[\sqrt{1-(L_x/\Lambda_y)^2}\frac{\pi}{2}\right].
\end{align}
Since the dependence of $\hat{\xi}$ on $\Lambda_y$ is complicated,
we again restrict the range of $\tilde{\Delta}'$ to the large $\Delta'$
regime,
\begin{align}
 \tilde{\Delta}'>\Delta'_c:=100/L_x\quad(\mbox{or }\Lambda_y/L_x>2.5),
\end{align}
as we have done in the linear theory.
Then, we can use the following approximations: 
\begin{align}
 \hat{\xi}(x)&\simeq-1
 -\frac{2}{\tilde{\Delta}' x}+O(x),\label{Taylor}\\
 \tilde{\Delta}' &\simeq16\frac{\Lambda_y^2}{L_x^3}.
\end{align}
The critical value $\Delta'_c=100/L_x$ is, of course, specific to the
equilibrium state \eqref{equilibrium}.

This external solution is matched to the inner solution by 
\begin{align}
-\sigma=\epsilon\hat{\xi}(\sigma)
=-\epsilon\left(1+\frac{2}{\tilde{\Delta}'\sigma}\right)+O(\epsilon\sigma),
\end{align}
which gives, by neglecting $O(\epsilon\sigma)$,
\begin{align}
 \sigma
=&\epsilon\frac{1+\sqrt{1+\frac{8}{\epsilon\tilde{\Delta}'}}}{2}.\label{matching}
\end{align}
Note that $\sigma$ is larger than
$\epsilon$ as illustrated in Fig.~\ref{Local_reconnection}. Since the displacement map is area-preserving, we determine
$l_y$ by the relation,
\begin{align}
 \sigma l_y=\epsilon\frac{\Lambda_y}{4}.\label{incompressibility}
\end{align}

\subsection{Energy balance}

In  linear  tearing mode theory,
the released magnetic energy via reconnection is estimated by
\begin{align}
\delta W
=&-\epsilon^2\frac{L_y}{2}\left.
\hat{\psi}\frac{d\hat{\psi}}{dx}\right|_{x=-a}^{x=a},
 \end{align}
in terms of the perturbed flux function $\hat{\psi}=-(d\psi^{(0)}/dx)\hat{\xi}=\tau_H^{-1}x\hat{\xi}$, 
where $2a(\ll L_x)$ is the width of the boundary layer at $x=0$ (see Appendix A of
Ref.~\cite{Hirota}). This is true if 
 the eigenfunction
$\hat{\psi}(x)$ is smoothed out and flattened within the layer $[-a,a]$
by some sort of nonideal MHD effects.

For the nonlinear phase in question, we simply replace $a$ by $\sigma$
(and $L_y$ by $\Lambda_y$) because the flux function $\psi$ is flattened
within $[-\sigma,\sigma]$ by the formation of a magnetic island.
This idea is similar to the finite-amplitude generalization of
$\Delta'$ which is made by White {\it et al.}~\cite{White} for the
purpose of introducing a saturation phase to Rutherford's theory. In
either case, the island width $2\sigma$ grows as far as $\hat{\psi}d\hat{\psi}/dx|_{x=-\sigma}^{x=+\sigma}>0$.
Using the Taylor expansion \eqref{Taylor} and the relation \eqref{incompressibility},
we obtain
\begin{align}
 \delta W
=-\epsilon^2\frac{\Lambda_y}{2}\hat{\psi}\frac{d\hat{\psi}}{dx}\big|_{x=-\sigma}^{x=\sigma}
=-4 l_y\tau_H^{-2}\sigma^3.\label{magnetic energy}
\end{align}
We remark that the magnetic energy in  the area
$[0,\sigma]\times[0,l_y]$ at the equilibrium state ($t=0$) is also  of the  order of
$l_y\tau_H^{-2}\sigma^3$. Since this area is mapped to the internal region of the
magnetic island after the displacement, we can expect a corresponding  decrease in the total magnetic
energy, which agrees with the estimation \eqref{magnetic energy}.

In order to satisfy the energy conservation $\delta K+\delta W=0$, the
kinetic energy is required to satisfy $\delta K\propto l_y\sigma^3$.
To be  concise, let us assume $l_y=$ const. a priori
because this assumption turns out to yield the desired scaling $\delta K\propto\sigma^3$ as follows.

Since the kinetic energy is mostly concentrated on the
down-stream side due to the outflow from the X-shaped vortex
layer,  we  use $\phi^{(d)}$ in \eqref{outflow} and the orderings $l_y\gg
l_x$ and $l_y\gg\sigma$ to estimate the kinetic energy in the
down-stream region as
\begin{align}
 \int_0^{l_y}dy\int_0^{\frac{l_x}{l_y}y}dx\frac{|\nabla\phi|^2}{2}
=&\left(
1+\frac{l_x^2}{6\sigma^2}
\right)\frac{\dot{\sigma}^2}{2}\frac{l_y^3}{2l_x}[1+O(l_x/l_y)+O(\sigma/l_y)]\nonumber\\
\simeq&\frac{\dot{\sigma}^2}{2}\frac{l_y^3}{2l_x},\label{kinetic energy0}
\end{align}
where we have neglected $l_x^2/6\sigma^2$ in the last expression 
since $l_x/\sigma=1-l_y/l_{y0}$ is now constant and less than unity.
The same estimate is more easily obtained as follows.
Consider the flow passing through the box
$[0,l_x]\times[0,l_y]$. Since the inflow velocity into the box
is at most $\dot{\sigma}$, the outflow velocity, say $\overline{v}_y$,
is roughly determined by the incompressibility condition,
\begin{align}
 \dot{\sigma}l_y=\bar{v}_yl_x,
\end{align}
where $\bar{v}_y\gg\dot{\sigma}$ owing to $l_y\gg l_x$. 
The kinetic energy density
$\bar{v}_y^2/2$ multiplied by the area $l_xl_y/2$ reproduces the same
estimate as \eqref{kinetic energy0}.

In fact, the outflow also exists over the area $[0,l_x]\times[l_y,L_y/2]$
in Fig.~\ref{Local_reconnection} and
there are eight equivalent areas in the whole domain according to the parity.
Therefore, a plausible estimate of the total kinetic energy is 
\begin{align}
 \delta
 K=8\frac{\bar{v}_y^2}{2}\left[\frac{l_xl_y}{2}+l_x\left(\frac{L_y}{2}-l_y\right)\right]
=2\frac{\dot{\sigma}^2l_y^2}{l_x}\left(L_y-l_y\right)
=2\frac{\sigma^3l_y^2}{l_{y0}^2\tau_H^2}\frac{L_y-l_y}{1-l_y/l_{y0}},
\end{align}
where the evolution equation \eqref{sigma} for $\sigma$ has been used.
Since this $\delta K$ is proportional to $\sigma^3$, we can impose the energy
conservation law $\delta K+\delta W=0$, which determines
$l_{y0}$ with respect to $l_y$,
\begin{align}
\frac{l_{y0}}{l_y}=\frac{1+\sqrt{2\frac{L_y}{l_y}-1}}{2}.
\end{align}
Given this $l_{y0}$, the rate of decrease in the magnetic energy
\begin{align}
 \frac{\partial_t(\delta W)}{\delta W}=\frac{3\sigma}{l_{y0}\tau_H}
\end{align}
indicates that the shorter the length $l_y$, the faster the magnetic
energy decreases. Thus, the local reconnection (i.e., the local X-shape)
develops faster than the global one.

\begin{figure}
\includegraphics[width=10cm]{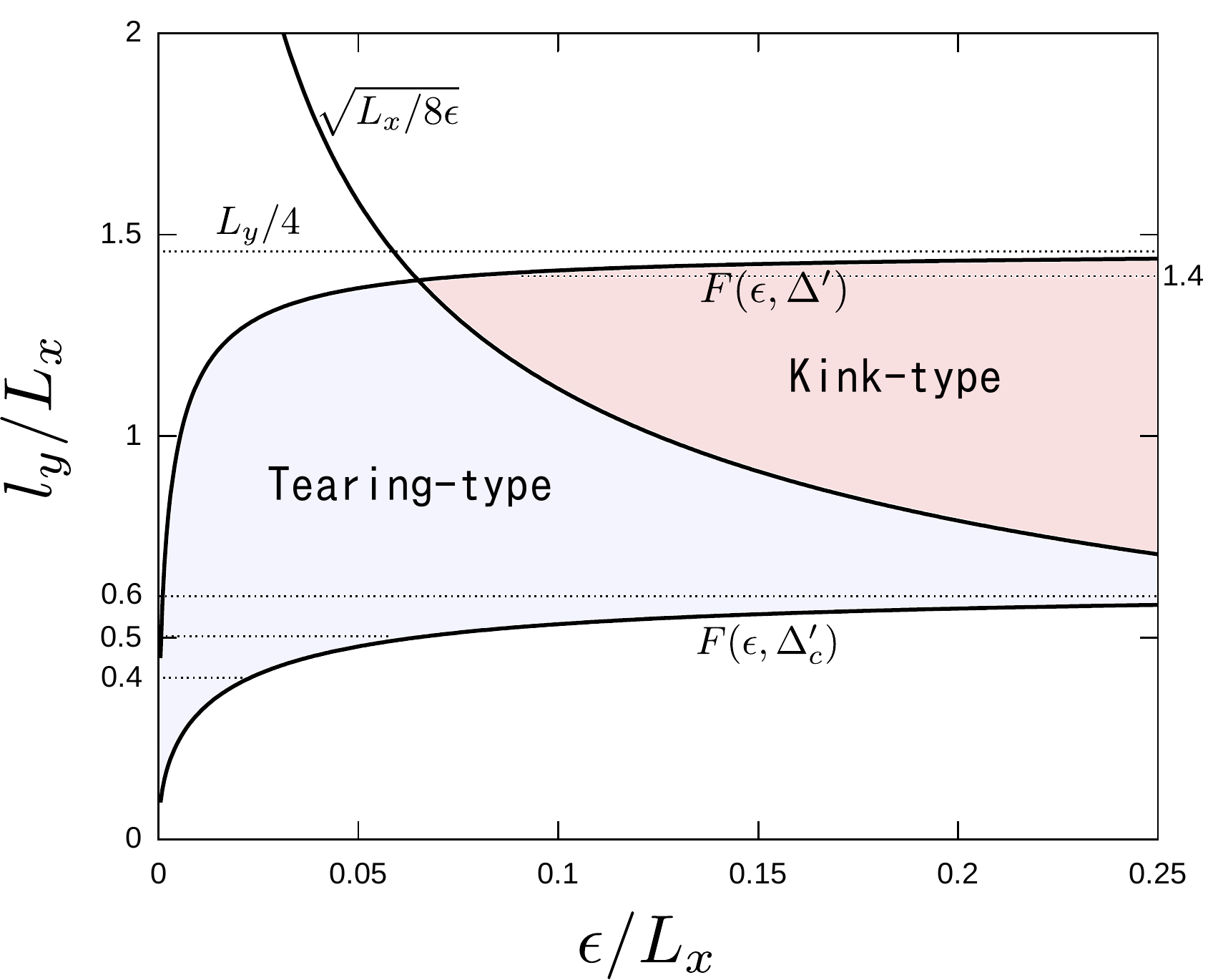} 
\caption{Range of $l_y$ corresponding to
 $\Delta_c'<\tilde{\Delta}'\le\Delta'$ for the case of $k_y/\alpha=L_x/L_y=0.171$.}\label{regime}
\end{figure}

However, there is a lower bound for $l_y$ since this argument is based
on the assumption $\Delta_c'<\tilde{\Delta}'\le\Delta'$, for which the
approximation $\tilde{\Delta}'=16\Lambda_y^2/L_x^3$ (and
$\Delta'=16L_y^2/L_x^3$) is valid.
Using \eqref{matching} and \eqref{incompressibility} with this approximation,
$l_y$ can be regarded as a function of $\epsilon$ and $\tilde{\Delta}'$,
\begin{align}
&\frac{l_y}{L_x}=\frac{1}{8}\frac{\sqrt{L_x\tilde{\Delta}'}}{1+\sqrt{1+\frac{8}{\epsilon\tilde{\Delta}'}}}=:F(\epsilon,\tilde{\Delta}'), \label{relation}
\end{align}
and hence $l_y/L_x$ should lie between $F(\epsilon,\Delta'_c)$ and $F(\epsilon,\Delta')$ as
shown in Fig.~\ref{regime}.

\subsection{Scaling of the explosive growth}

By rewriting \eqref{relation} as
\begin{align}
\left(1+\sqrt{1+\frac{8}{\epsilon\tilde{\Delta}'}}\right)\sqrt{\frac{8}{\epsilon\tilde{\Delta}'}}
=\frac{L_x}{l_y}\sqrt{\frac{L_x}{8\epsilon}},\label{relation2}
\end{align}
this relation is found to have two kinds
of scaling   depending on whether its right hand side is much smaller or larger than  unity, 

First, when $l_y/L_x\gg\sqrt{L_x/8\epsilon}$, the relation reduces to
\begin{align}
\frac{8}{\epsilon\tilde{\Delta}'}
=\frac{L_x^3}{32l_y^2\epsilon}\ll1.
\end{align}
Since $\sigma\simeq\epsilon$ in this case, we obtain the same explosive growth as \eqref{sigma},
\begin{align}
 \dot{\epsilon}=\frac{\epsilon^2}{\tau_Hl_{y0}},\label{kink-type}
\end{align}
in terms of the displacement $\epsilon$ at $x=L_x/4$.
We refer this scaling as  kink-type because
$\tilde{\Delta}'$ is so large that the external solution is similar to the
kink mode ($\sigma\simeq\epsilon$, $l_y\simeq\Lambda_y/4$). 

On the other hand, when $l_y/L_x\ll\sqrt{L_x/8\epsilon}$, the relation
\eqref{relation2} reduces to
\begin{align}
\frac{8}{\epsilon\tilde{\Delta}'}
=\sqrt{\frac{L_x^3}{8l_y^2\epsilon}}\gg1.
\end{align}
By noting that
\begin{align}
 \sigma\simeq\frac{\epsilon}{2}\sqrt{\frac{8}{\epsilon\tilde{\Delta}'}}
=\frac{\epsilon^{3/4}}{2}\left(\frac{L_x^3}{8l_y^2}\right)^{1/4},
\end{align}
the explosive growth \eqref{sigma} becomes
\begin{align}
 \dot{\epsilon}=\frac{\epsilon^{7/4}}{\tau_Hl_{y0}}
\frac{2}{3}\left(\frac{L_x^3}{8l_y^2}\right)^{1/4}.\label{tearing-type}
\end{align}
We refer this scaling as  tearing-type because $\tilde{\Delta}'$ is so small that the external solution is similar to the
tearing mode ($\sigma>\epsilon$, $l_y<\Lambda_y/4$). 

\begin{figure}
\includegraphics[width=10cm]{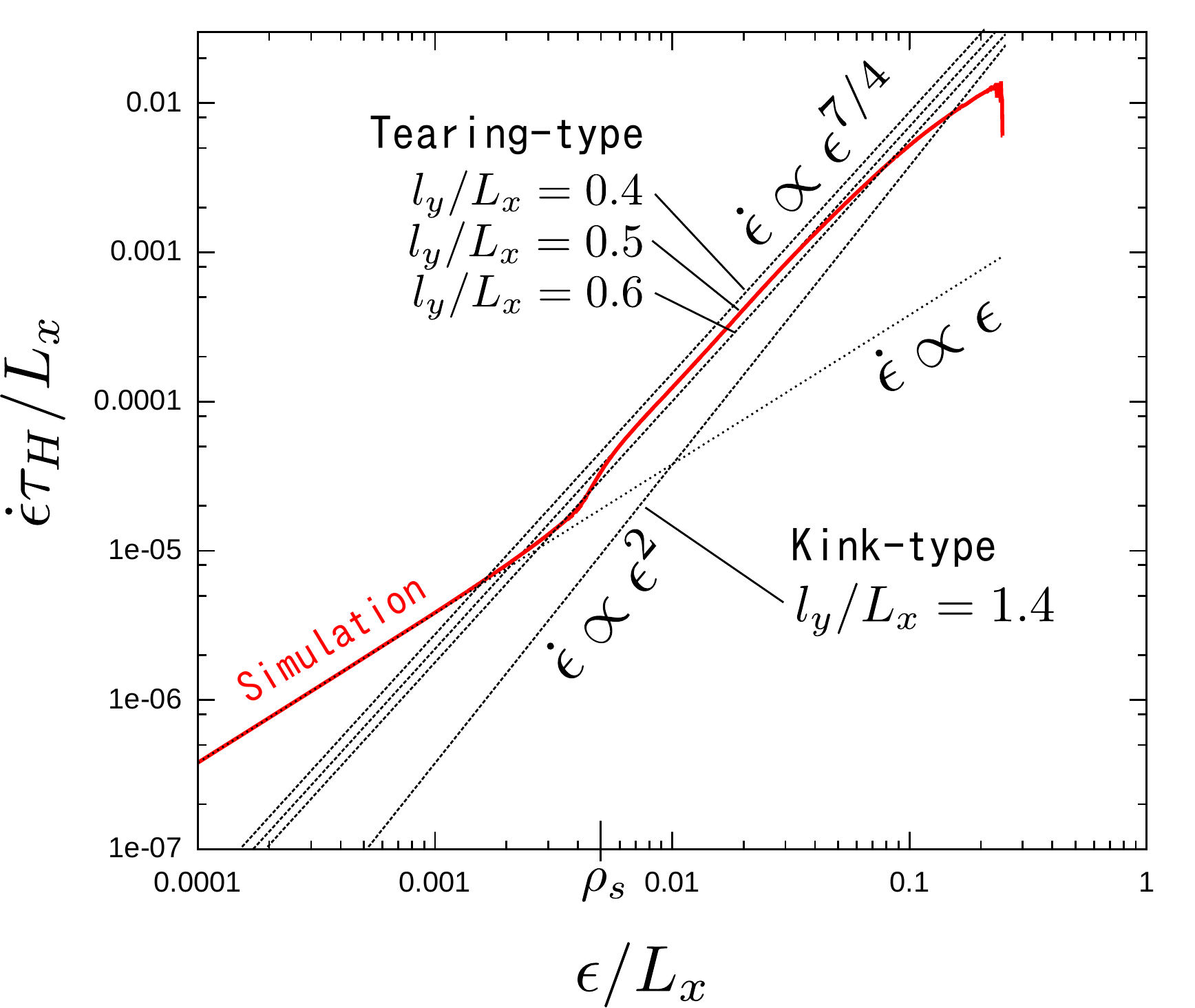} 
\caption{Logarithmic plot of the displacement $\epsilon$ versus its
 time-derivative $\dot{\epsilon}$ for $d_e=\rho_s=0.005L_x$,
 $k_y/\alpha=0.171$. Dotted lines are theoretically derived from the kink-type scaling
 \eqref{kink-type} with $l_y/L_x=1.4$ and the tearing-type scaling
 \eqref{tearing-type} with $l_y/L_x=0.4,0.5,0.6$.}\label{growth}
\end{figure}

The boundary line $l_y/L_x=\sqrt{L_x/8\epsilon}$ between the kink-type
and tearing-type regimes is also drawn in Fig.~\ref{regime}.
Since the magnetic energy is
released more effectively for the smaller $l_y$, the fastest
reconnection occurs near the lower bound $l_y/L_x=0.4\sim0.6$.
Figure~\ref{growth} shows that the tearing-type scaling
\eqref{tearing-type} for $l_y/L_x=0.4,0.5$, and $0.6$ agrees
well with the simulation results. For comparison, we also draw
the kink-type scaling \eqref{kink-type} with $l_y=1.4L_x\simeq L_y/4$
as a global reconnection model, which is indeed slower than the
simulation results.
We can confirm that the stream lines in Fig.~\ref{contours3}(a) are more like the tearing-type
($\sigma>\epsilon$, $l_y<\Lambda_y/4$).
Although the current layers in Fig.~\ref{contours3}(c) are actually
curved, they are locally regarded as straight lines around the origin as
in Fig.~\ref{Local_reconnection} and seem to have the length $l_y/L_x\simeq0.5$.
Note that the same $l_y/L_x\simeq0.5$ is also observed in
Fig.~\ref{growth_aspect}(b) and (c) since this $l_y/L_x$ is determined
independently of $L_y$ and $d_e=\rho_s$ as shown in \eqref{relation}.

\section{Summary}\label{sec:summary}

We have investigated the nonlinear evolution of a collisionless tearing
mode that can grow explosively with the formation of an X-shaped
current-vortex layer due to the coexistence of electron inertia and
temperature effects, where we have assumed $d_e=\rho_s$ for simplicity.

For the equilibrium state given in \eqref{equilibrium} and the wavenumber
 $k_y=2\pi/L_y$, the tearing mode is linearly unstable when the tearing
 index $\Delta'$ (which is a function of $L_y/L_x$) is positive.
The simulation results show that   explosive growth occurs when
 $\Delta'\gtrsim0.65/\rho_s$. More specifically, the amplitude $\epsilon$ of the displacement 
 at $x=\pm L_x/4$ exceeds $\rho_s$ and then grows explosively; $\dot{\epsilon}\propto\epsilon^n$, $n>1$.
By observing this explosive phase in detail for $d_e=\rho_s<0.01$, we find that the X-shaped layer widens locally
 around the reconnection point and its length scale ($\simeq2l_y$) seems to be
 unrelated to the wavelength $L_y$ (and $\Delta'$ as well) of the linear eigenmode.

To explain this locally enhanced reconnection, we have developed a theoretical model
in which the magnetic flux $\psi$ is assumed to be conserved (like ideal
MHD) except  within  the thin X-shaped layer. Namely, the two-fluid conservation laws
\eqref{conservation1}, \eqref{conservation2} are invoked only within
the layer to obtain the matching conditions across it.
The external solution is approximated by a linear tearing eigenmode
that has a shorter wavenumber $\Lambda_y$ than $L_y$ and a
smaller tearing index $\tilde{\Delta}'$ than $\Delta'$. We have restricted
our consideration to the range
$\Delta_c'(=100/L_x)<\tilde{\Delta}'\le\Delta'$ (or $2.5L_x<\Lambda_y<L_y$),
in which a simple expression $\tilde{\Delta}'=16\Lambda_y^2/L_x^3$ holds
and the length of the local X-shape ($\simeq2l_y$) is related to $\tilde{\Delta}'$ by
\eqref{relation}. As shown in
Fig.~\ref{regime},
 we have found that there are two kinds of scaling depending on
whether the external solution is kink-type or tearing-type.
The faster reconnection is theoretically predicted at the shorter $l_y$, 
namely, at the lower bound of this range, $l_y\simeq0.5L_x$,
$\tilde{\Delta}'\simeq\Delta_c'$ and $\Lambda_y\simeq2.5L_x$, which
belongs to the
tearing-type regime. The simulation results indeed agree with the
tearing-type scaling with  the explosive growth rate
$\dot{\epsilon}\propto\epsilon^{7/4}$ and they corroborate other properties predicted by
this local reconnection model.

In comparison with the classical Petschek reconnection model~\cite{Petschek} in which the X-shaped
boundary layer is composed of stationary slow-mode shocks, our model suggests
that the X-shaped current-vortex layer is kinematically generated by
ideal, incompressible and accelerated fluid motion in accordance with the
two-fluid conservation laws and the energy conservation. The explosive
growth rate \eqref{tearing-type} with $l_y\simeq0.5L_x$ is moreover independent of the
microscopic scale $d_e=\rho_s$ and hence reaches the Alfv\'en speed
$\dot{\epsilon}\sim L_x^2/(\tau_H l_{y0})$ at the fully reconnected stage $\epsilon\sim L_x/4$.
This is faster than the explosive growth $\dot{\epsilon}\sim
k_yd_e^{1/2}\epsilon^{3/2}/\tau_H$ that is
caused by the Y-shaped layer
in the presence of only electron inertia (see our  previous work~\cite{Hirota}).

The two-field equations \eqref{vorticity} and \eqref{Ohm} can be derived from
  gyrokinetic and gyro-fluid  equations by taking the fluid moments and
then neglecting  the ion  pressure and electron and ion  gyroradii~\cite{Ishizawa,Comisso}.
This fact suggests our  present results are  a barebones model for  fast reconnection, but further generalizations including the  case $d_e\ne\rho_s$  are suggested for  future work. 
The existence of more than one microscopic scale gives rise to nested
boundary layers,  as already known from  the linear analysis.  
The nonlinear evolution of such   nested boundary layers would
be more complicated than that of the single boundary layer ($d_e=\rho_s$)
we have discussed.
Nevertheless, if the outermost layer is sufficiently thinner than the
island width and the energy balance is dominated by ideal MHD, we expect
a similar X-shaped layer and explosive growth, 
since it  is unlikely that any other structure can  exist that is more efficient for releasing  magnetic energy.
Unfortunately,   present computational resources are  not
enough to observe the explosive phase for a sufficiently long period in
the presence of the nested bounded layers.  For example, when $d_e\ll\rho_s$,   linear analysis indicates that the
innermost layer width $\sim d_e^{4/3}\rho_s^{-1/3}$ is
 even narrower than $d_e$ and demands more computational grids.
Further  advancements in computational performance and technique will be
essential for studies of explosive reconnections in more general
collisionless plasma models.

\begin{acknowledgments}
    The authors are grateful to Dr.~Masatoshi Yagi and Dr.~Yasutomo Ishii for useful discussions and suggestions.
MH and YH were  supported by JSPS KAKENHI Grant Number~25800308.  PJM was  supported by U.S. Dept.\ of Energy Contract \# DE-FG02-04ER54742.

\end{acknowledgments}


%
%


\begin{thebibliography}{9}
 \addcontentsline{toc}{chapter}{\bibname}
 \markboth{\bibname}{}
 \bibitem{Sweet}
	 P.A. Sweet,
	 {\it Electromagnetic Phenomena in Cosmical Physics},
	 IAU Symp. No. 6, edited by B. Lehnert (Cambridge Press, London,
	 1958). P. 123;
	 E.N. Parker,
	 J. Geophys. Res. {\bf 62}, 509 (1957).
 \bibitem{Petschek}
	 H.E. Petschek,
	 {\it Physics of Solar Flares}. Edited by W. N. Hess (NASA SP-50, Washington DC, 1964), p. 425.

 \bibitem{Ugai}
	 M. Ugai and T. Tsuda,
	 J. Plasma Phys. {\bf 17}, 337 (1977).

 \bibitem{Hazeltine}
	 R.D. Hazeltine, M. Kotschenreuther and P. J. Morrison,
	 Phys. Fluids {\bf 28}, 2466 (1985).
 \bibitem{Schep}
	 T. J. Schep, F. Pegoraro and B. N. Kuvshinov,
	 Phys. Plasmas, {\bf 1}, 2843 (1994).
 \bibitem{Kuvshinov}
         B. N. Kuvshinov, F. Pegoraro and T. J. Schep, 
         Phys. Lett. A, {\bf 191}, 296 (1994).
 \bibitem{Fitzpatrick}
	 R. Fitzpatrick and F. Porcelli,
	 Phys. Plasmas, {\bf 11} 4713 (2004).

 \bibitem{Snyder}
	 P. B. Snyder and G. W. Hammett,
	 Phys. Plasmas, {\bf 8}, 3199 (2001).
 \bibitem{Waelbroeck}
	 F. L. Waelbroeck and E. Tassi,
	 Commun. Nonlinear Sci. Numer. Simul. {\bf 17}, 2171 (2012).

 \bibitem{Frieman}
	 E. A. Frieman and L. Chen,
	 Phys. Fluids, {\bf 25}, 502 (1982).
 \bibitem{Zocco}
	 A. Zocco and A. A. Schekochihin,
	 {\it Phys. Plasmas} {\bf 18} 102309 (2011).

 \bibitem{Aydemir}
	 A. Y. Aydemir,
	 Phys. Fluids B, {\bf 4}, 2469 (1992).
 \bibitem{Ottaviani}
	 M. Ottaviani and F. Porcelli, 
	 Phys. Rev. Lett., {\bf 71}, 3802 (1993).
\bibitem{Cafaro}
	 E. Cafaro, D. Grasso, F. Pegoraro, F. Porcelli and A. Saluzzi,
	Phys. Rev. Lett., {\bf 80}, 4430 (1998).
 \bibitem{Bhattacharjee}
	 A. Bhattacharjee, K. Germaschewski and C. S. Ng
	 Phys. Plasmas, {\bf 12}, 042305 (2005).
 \bibitem{Matsumoto}
	 T. Matsumoto, H. Naitou, S. Tokuda and Y. Kishimoto,
	 Phys. Plasmas {\bf 12}, 092505 (2005)
 \bibitem{Biancalani}
	 A. Biancalani and B. D. Scott,
	 Europhys. Lett. {\bf 97}, 15005 (2012).
 \bibitem{Comisso}
	 L. Comisso, D. Grasso, F. L. Waelbroeck and D. Borgogno,
	 Phys. Plasmas {\bf 20}, 092118 (2013).
 \bibitem{Ishizawa}
	 A. Ishizawa and T.-H. Watanabe,
	 Phys. Plasmas {\bf 20}, 102116 (2013).

 \bibitem{Drake}
	 J. F. Drake,
	 Phys. Fluids, {\bf 21}, 1777 (1978).
 \bibitem{Basu}
	 B. Basu and B. Coppi,
	 Phys. Fluids, {\bf 24}, 465 (1981).
 \bibitem{Porcelli}
	 F. Porcelli,
	 Phys. Rev. Lett., {\bf 66}, 425 (1991).








 
 \bibitem{Rutherford}
	 P. H. Rutherford
	 Phys. Fluids, {\bf 16}, 1903 (1973).
 \bibitem{White}
	 R. B. White, D. A. Monticello, M. N. Rosenbluth and
	 B. V. Waddell,
	 Phys. Fluids, {\bf 20}, 800 (1977).

 \bibitem{Hirota}
	 M. Hirota, P.J. Morrison, Y. Ishii, M. Yagi and N. Aiba,
	 Nucl. Fusion, {\bf 53}, 063024 (2013).









 \bibitem{Hazeltine2}
	 R. D. Hazeltine, C. T. Hsu and P. J. Morrison,
	 Phys. Fluids {\bf 30}, 3204 (1987).

\bibitem{morrison-greene}
	 Morrison  P.J.  and J. M. Greene 1980 {\it Phys. Rev. Letts.} {\bf 45}  790

\bibitem{morrison98}
	 Morrison  P.J. 1998 {\it Rev. Mod. Phys.} {\bf 70}  467  
	 

 \bibitem{Kuvshinov2}
	 B. N. Kuvshinov, V. P. Lakhin, F. Pegoraro and T. J. Schep,
	 J. Plasma Physics, {\bf 59}, 727  (1998)
 \bibitem{Grasso}
	 D. Grasso, F. Califano, F. Pegoraro and F. Porcelli,
	 Plasma Phys. Control. Fusion, {\bf 41}, 1497 (1999).
 \bibitem{Tassi}
	 E. Tassi, P. J.  Morrison, D. Grasso and F. Pegoraro
	 Nucl. Fusion, {\bf 50}, 034007 (2010).

 \bibitem{Grasso2}
	 D. Grasso, F. Califano, F. Pegoraro and F. Porcelli,
	 Phys. Rev. Lett. {\bf 86}, 5051 (2001).



 \bibitem{Biskamp2}
	 D. Biskamp,
	 {\it Magnetic Reconnection in Plasmas} (Cambridge
	 University Press, Cambridge, 2000)

\end{thebibliography}
\end{document}